\documentclass[12pt,english]{article}
\usepackage{avant}
\usepackage{xr}
\usepackage[T1]{fontenc}
\usepackage[latin9]{inputenc}
\usepackage{geometry}
\usepackage{tikz}
\usepackage{tkz-graph}
\tikzstyle{vertex}=[circle,fill=white,draw]
\usepackage{amsmath}
\usepackage{amssymb}
\usepackage{stmaryrd}
\usetikzlibrary{arrows.meta,calc,decorations.markings,math,arrows.meta}
\usepackage{babel}
\usepackage{array}
\usepackage{float}
\usepackage{multirow}
\usepackage{amsmath}
\usepackage{amssymb}
\usepackage{stmaryrd}
\usepackage[authoryear]{natbib}
\usepackage[unicode=true,
 bookmarks=false,
 breaklinks=false,pdfborder={0 0 1},colorlinks=false]
 {hyperref}
\usepackage{breakurl}

\makeatletter

\floatstyle{ruled}
\newfloat{algorithm}{tbp}{loa}
\providecommand{\algorithmname}{Algorithm}
\floatname{algorithm}{\protect\algorithmname}

\usepackage{amsthm}\usepackage{bm}\usepackage{graphicx}\usepackage{rotating}\usepackage{array}\usepackage{booktabs}\usepackage{color}\usepackage{algorithmic}
\usepackage{algorithm}\usepackage{multirow}\usepackage{enumerate}\usepackage{graphicx}\usepackage{caption}\usepackage{subcaption}

\floatstyle{ruled}
\newfloat{algorithm}{tbp}{loa}
\providecommand{\algorithmname}{Algorithm}
\floatname{algorithm}{\protect\algorithmname}

\usepackage{vmargin}
\setmarginsrb{1.0in}{0.5in}{1.0in}{1.5in}{0.5in}{0.5in}{0.5in}{0.5in}

\@ifundefined{definecolor}{color}{}
\RequirePackage[displaymath]{lineno}

\@ifundefined{definecolor}{color}{}
\usepackage{amsfonts}\usepackage{latexsym}

\usepackage{mathrsfs}\usepackage{amsthm}\usepackage{latexsym}\usepackage{booktabs}\@ifundefined{definecolor}{color}{}

\usepackage{times}\@ifundefined{definecolor}{color}{}

\numberwithin{equation}{section}
\numberwithin{figure}{section}

\newcommand{\T}{^{\mbox{\tiny {\sf T}}}}

\usepackage{babel}

\makeatother

\begin{document}
{\setlength{\baselineskip}{1.45\baselineskip}

\global\long\def\mba{\mathbf{a}}
 \global\long\def\mbA{\mathbf{A}}
 \global\long\def\mbb{\mathbf{b}}
 \global\long\def\mbB{\mathbf{B}}
 \global\long\def\mbc{\mathbf{c}}
 \global\long\def\mbC{\mathbf{C}}
 \global\long\def\mbd{\mathbf{d}}
 \global\long\def\mbD{\mathbf{D}}
 \global\long\def\mbe{\mathbf{e}}
 \global\long\def\mbE{\mathbf{E}}
 \global\long\def\mbf{\mathbf{f}}
 \global\long\def\mbF{\mathbf{F}}
 \global\long\def\mbg{\mathbf{g}}
 \global\long\def\mbG{\mathbf{G}}
 \global\long\def\mbh{\mathbf{h}}
 \global\long\def\mbH{\mathbf{H}}
 \global\long\def\mbi{\mathbf{i}}
 \global\long\def\mbI{\mathbf{I}}
 \global\long\def\mbj{\mathbf{j}}
 \global\long\def\mbJ{\mathbf{J}}
 \global\long\def\mbk{\mathbf{k}}
 \global\long\def\mbK{\mathbf{K}}
 \global\long\def\mbl{\mathbf{l}}
 \global\long\def\mbL{\mathbf{L}}
 \global\long\def\mbm{\mathbf{m}}
 \global\long\def\mbM{\mathbf{M}}
 \global\long\def\mbn{\mathbf{n}}
 \global\long\def\mbN{\mathbf{N}}
 \global\long\def\mbo{\mathbf{o}}
 \global\long\def\mbO{\mathbf{O}}
 \global\long\def\mbp{\mathbf{p}}
 \global\long\def\mbP{\mathbf{P}}
 \global\long\def\mbq{\mathbf{q}}
 \global\long\def\mbQ{\mathbf{Q}}
 \global\long\def\mbr{\mathbf{r}}
 \global\long\def\mbR{\mathbf{R}}
 \global\long\def\mbs{\mathbf{s}}
 \global\long\def\mbS{\mathbf{S}}
 \global\long\def\mbt{\mathbf{t}}
 \global\long\def\mbT{\mathbf{T}}
 \global\long\def\mbu{\mathbf{u}}
 \global\long\def\mbU{\mathbf{U}}
 \global\long\def\mbv{\mathbf{v}}
 \global\long\def\mbV{\mathbf{V}}
 \global\long\def\mbw{\mathbf{w}}
 \global\long\def\mbW{\mathbf{W}}
 \global\long\def\mbx{\mathbf{x}}
 \global\long\def\mbX{\mathbf{X}}
 \global\long\def\mby{\mathbf{y}}
 \global\long\def\mbY{\mathbf{Y}}
 \global\long\def\mbz{\mathbf{z}}
 \global\long\def\mbZ{\mathbf{Z}}
 \global\long\def\barmbZ{\bar{\mathbf{Z}}}

\global\long\def\hatmba{\widehat{\mathbf{a}}}
 \global\long\def\hatmbA{\widehat{\mathbf{A}}}
 \global\long\def\hatmbb{\widehat{\mathbf{b}}}
 \global\long\def\hatmbB{\widehat{\mathbf{B}}}
 \global\long\def\hatmbc{\widehat{\mathbf{c}}}
 \global\long\def\hatmbC{\widehat{\mathbf{C}}}
 \global\long\def\hatmbd{\widehat{\mathbf{d}}}
 \global\long\def\hatmbD{\widehat{\mathbf{D}}}
 \global\long\def\hatmbe{\widehat{\mathbf{e}}}
 \global\long\def\hatmbE{\widehat{\mathbf{E}}}
 \global\long\def\hatmbf{\widehat{\mathbf{f}}}
 \global\long\def\hatmbF{\widehat{\mathbf{F}}}
 \global\long\def\hatmbg{\widehat{\mathbf{g}}}
 \global\long\def\hatmbG{\widehat{\mathbf{G}}}
 \global\long\def\hatmbh{\widehat{\mathbf{h}}}
 \global\long\def\hatmbH{\widehat{\mathbf{H}}}
 \global\long\def\hatmbi{\widehat{\mathbf{i}}}
 \global\long\def\hatmbI{\widehat{\mathbf{I}}}
 \global\long\def\hatmbj{\widehat{\mathbf{j}}}
 \global\long\def\hatmbJ{\widehat{\mathbf{J}}}
 \global\long\def\hatmbk{\widehat{\mathbf{k}}}
 \global\long\def\hatmbK{\widehat{\mathbf{K}}}
 \global\long\def\hatmbl{\widehat{\mathbf{l}}}
 \global\long\def\hatmbL{\widehat{\mathbf{L}}}
 \global\long\def\hatmbm{\widehat{\mathbf{m}}}
 \global\long\def\hatmbM{\widehat{\mathbf{M}}}
 \global\long\def\hatmbn{\widehat{\mathbf{n}}}
 \global\long\def\hatmbN{\widehat{\mathbf{N}}}
 \global\long\def\hatmbo{\widehat{\mathbf{o}}}
 \global\long\def\hatmbO{\widehat{\mathbf{O}}}
 \global\long\def\hatmbp{\widehat{\mathbf{p}}}
 \global\long\def\hatmbP{\widehat{\mathbf{P}}}
 \global\long\def\hatmbq{\widehat{\mathbf{q}}}
 \global\long\def\hatmbQ{\widehat{\mathbf{Q}}}
 \global\long\def\hatmbr{\widehat{\mathbf{r}}}
 \global\long\def\hatmbR{\widehat{\mathbf{R}}}
 \global\long\def\hatmbs{\widehat{\mathbf{s}}}
 \global\long\def\hatmbS{\widehat{\mathbf{S}}}
 \global\long\def\hatmbt{\widehat{\mathbf{t}}}
 \global\long\def\hatmbT{\widehat{\mathbf{T}}}
 \global\long\def\hatmbu{\widehat{\mathbf{u}}}
 \global\long\def\hatmbU{\widehat{\mathbf{U}}}
 \global\long\def\hatmbv{\widehat{\mathbf{v}}}
 \global\long\def\hatmbV{\widehat{\mathbf{V}}}
 \global\long\def\hatmbw{\widehat{\mathbf{w}}}
 \global\long\def\hatmbW{\widehat{\mathbf{W}}}
 \global\long\def\hatmbx{\widehat{\mathbf{x}}}
 \global\long\def\hatmbX{\widehat{\mathbf{X}}}
 \global\long\def\hatmby{\widehat{\mathbf{y}}}
 \global\long\def\hatmbY{\widehat{\mathbf{Y}}}
 \global\long\def\hatmbz{\widehat{\mathbf{z}}}
 \global\long\def\hatmbZ{\widehat{\mathbf{Z}}}

\global\long\def\tilmba{\widetilde{\mathbf{a}}}
 \global\long\def\tilmbA{\widetilde{\mathbf{A}}}
 \global\long\def\tilmbb{\widetilde{\mathbf{b}}}
 \global\long\def\tilmbB{\widetilde{\mathbf{B}}}
 \global\long\def\tilmbc{\widetilde{\mathbf{c}}}
 \global\long\def\tilmbC{\widetilde{\mathbf{C}}}
 \global\long\def\tilmbd{\widetilde{\mathbf{d}}}
 \global\long\def\tilmbD{\widetilde{\mathbf{D}}}
 \global\long\def\tilmbe{\widetilde{\mathbf{e}}}
 \global\long\def\tilmbE{\widetilde{\mathbf{E}}}
 \global\long\def\tilmbf{\widetilde{\mathbf{f}}}
 \global\long\def\tilmbF{\widetilde{\mathbf{F}}}
 \global\long\def\tilmbg{\widetilde{\mathbf{g}}}
 \global\long\def\tilmbG{\widetilde{\mathbf{G}}}
 \global\long\def\tilmbh{\widetilde{\mathbf{h}}}
 \global\long\def\tilmbH{\widetilde{\mathbf{H}}}
 \global\long\def\tilmbi{\widetilde{\mathbf{i}}}
 \global\long\def\tilmbI{\widetilde{\mathbf{I}}}
 \global\long\def\tilmbj{\widetilde{\mathbf{j}}}
 \global\long\def\tilmbJ{\widetilde{\mathbf{J}}}
 \global\long\def\tilmbk{\widetilde{\mathbf{k}}}
 \global\long\def\tilmbK{\widetilde{\mathbf{K}}}
 \global\long\def\tilmbl{\widetilde{\mathbf{l}}}
 \global\long\def\tilmbL{\widetilde{\mathbf{L}}}
 \global\long\def\tilmbm{\widetilde{\mathbf{m}}}
 \global\long\def\tilmbM{\widetilde{\mathbf{M}}}
 \global\long\def\tilmbn{\widetilde{\mathbf{n}}}
 \global\long\def\tilmbN{\widetilde{\mathbf{N}}}
 \global\long\def\tilmbo{\widetilde{\mathbf{o}}}
 \global\long\def\tilmbO{\widetilde{\mathbf{O}}}
 \global\long\def\tilmbp{\widetilde{\mathbf{p}}}
 \global\long\def\tilmbP{\widetilde{\mathbf{P}}}
 \global\long\def\tilmbq{\widetilde{\mathbf{q}}}
 \global\long\def\tilmbQ{\widetilde{\mathbf{Q}}}
 \global\long\def\tilmbr{\widetilde{\mathbf{r}}}
 \global\long\def\tilmbR{\widetilde{\mathbf{R}}}
 \global\long\def\tilmbs{\widetilde{\mathbf{s}}}
 \global\long\def\tilmbS{\widetilde{\mathbf{S}}}
 \global\long\def\tilmbt{\widetilde{\mathbf{t}}}
 \global\long\def\tilmbT{\widetilde{\mathbf{T}}}
 \global\long\def\tilmbu{\widetilde{\mathbf{u}}}
 \global\long\def\tilmbU{\widetilde{\mathbf{U}}}
 \global\long\def\tilmbv{\widetilde{\mathbf{v}}}
 \global\long\def\tilmbV{\widetilde{\mathbf{V}}}
 \global\long\def\tilmbw{\widetilde{\mathbf{w}}}
 \global\long\def\tilmbW{\widetilde{\mathbf{W}}}
 \global\long\def\tilmbx{\widetilde{\mathbf{x}}}
 \global\long\def\tilmbX{\widetilde{\mathbf{X}}}
 \global\long\def\tilmby{\widetilde{\mathbf{y}}}
 \global\long\def\tilmbY{\widetilde{\mathbf{Y}}}
 \global\long\def\tilmbz{\widetilde{\mathbf{z}}}
 \global\long\def\tilmbZ{\widetilde{\mathbf{Z}}}
 
\renewcommand{\Vec}{\mathrm{Vec}}
\newcommand{\bSigma}{{\bm \Sigma}}
\newcommand{\bzero}{\mathbf{0}}

\newcommand{\bX}{\mathbf{X}}
\newcommand{\bY}{\mathbf{Y}}
\newcommand{\bS}{\mathbf{S}}
\newcommand{\bx}{\mathbf{x}}
\newcommand{\bV}{\mathbf{V}}
\newcommand{\bv}{\mathbf{v}}
\newcommand{\bG}{\mathbf{G}}
\newcommand{\bbeta}{{\bm\beta}}
\newcommand{\bgamma}{{\bm\gamma}}
\newcommand{\bGamma}{{\bm\Gamma}}
\newcommand{\bLambda}{{\bm\Lambda}}

\newcommand{\bA}{\mathbf{A}}
\newcommand{\bB}{\mathbf{B}}
\newcommand{\bC}{\mathbf{C}}
\newcommand{\bD}{\mathbf{D}}
\newcommand{\bI}{\mathbf{I}}
\newcommand{\bJ}{\mathbf{J}}
\newcommand{\bR}{\mathbf{R}}
\newcommand{\bu}{\mathbf{u}}
\newcommand{\bz}{\mathbf{z}}
\newcommand{\f}{\mathbf{f}}
\newcommand{\bzeta}{{\bm \zeta}}
\newcommand{\btheta}{{\bm \theta}}
\newcommand{\bP}{\mathbf{P}}
\newcommand{\bT}{\mathbf{T}}
\newcommand{\bZ}{\mathbf{Z}}
\newcommand{\bepsilon}{\bm \epsilon}
\newcommand{\bmu}{\bm \mu}
\newcommand{\bxi}{\bm \xi}
\newcommand{\bW}{\mathbf{W}}
\newcommand{\bOmega}{\bm \Omega}
\newcommand{\bU}{\mathbf{U}}
\newcommand{\bTheta}{\bm \Theta}
\newcommand{\bpi}{\bm \pi}
\newcommand{\balpha}{\bm \alpha}
\newcommand{\bdelta}{\bm \delta}
\newcommand{\bF}{\mathbf{F}}

 \global\long\def\hatf{\widehat{f}}

\global\long\def\bolalpha{\boldsymbol{\alpha}}
 \global\long\def\bolbeta{\boldsymbol{\beta}}
 \global\long\def\bolgamma{\boldsymbol{\gamma}}
 \global\long\def\boldelta{\boldsymbol{\delta}}
 \global\long\def\bolepsilon{\boldsymbol{\epsilon}}
 \global\long\def\bolzeta{\boldsymbol{\zeta}}
 \global\long\def\boleta{\boldsymbol{\eta}}
 \global\long\def\boltheta{\boldsymbol{\theta}}
 \global\long\def\bolkappa{\boldsymbol{\kappa}}
 \global\long\def\bollambda{\boldsymbol{\lambda}}
 \global\long\def\bolmu{\boldsymbol{\mu}}
 \global\long\def\bolnu{\boldsymbol{\nu}}
 \global\long\def\bolxi{\boldsymbol{\xi}}
 \global\long\def\bolpi{\boldsymbol{\pi}}
 \global\long\def\bolrho{\boldsymbol{\rho}}
 \global\long\def\bolsigma{\boldsymbol{\sigma}}
 \global\long\def\boltau{\boldsymbol{\tau}}
 \global\long\def\bolphi{\boldsymbol{\phi}}
 \global\long\def\bolchi{\boldsymbol{\chi}}
 \global\long\def\bolpsi{\boldsymbol{\psi}}
 \global\long\def\bolomega{\boldsymbol{\omega}}
 \global\long\def\bolGamma{\boldsymbol{\Gamma}}
 \global\long\def\bolDelta{\boldsymbol{\Delta}}
 \global\long\def\bolTheta{\boldsymbol{\Theta}}
 \global\long\def\bolLambda{\boldsymbol{\Lambda}}
 \global\long\def\bolPi{\boldsymbol{\Pi}}
 \global\long\def\bolSigma{\boldsymbol{\Sigma}}
 \global\long\def\bolPhi{\boldsymbol{\Phi}}
 \global\long\def\bolPsi{\boldsymbol{\Psi}}
 \global\long\def\bolOmega{\boldsymbol{\Omega}}

\global\long\def\hatbolalpha{\widehat{\boldsymbol{\alpha}}}
 \global\long\def\hatbolbeta{\widehat{\boldsymbol{\beta}}}
 \global\long\def\hatbolgamma{\widehat{\boldsymbol{\gamma}}}
 \global\long\def\hatboldelta{\widehat{\boldsymbol{\delta}}}
 \global\long\def\hatbolepsilon{\widehat{\boldsymbol{\epsilon}}}
 \global\long\def\hatbolzeta{\widehat{\boldsymbol{\zeta}}}
 \global\long\def\hatboleta{\widehat{\boldsymbol{\eta}}}
 \global\long\def\hatboltheta{\widehat{\boldsymbol{\theta}}}
 \global\long\def\hatbolkappa{\widehat{\boldsymbol{\kappa}}}
 \global\long\def\hatbollambda{\widehat{\boldsymbol{\lambda}}}
 \global\long\def\hatbolmu{\widehat{\boldsymbol{\mu}}}
 \global\long\def\hatbolnu{\widehat{\boldsymbol{\nu}}}
 \global\long\def\hatbolxi{\widehat{\boldsymbol{\xi}}}
 \global\long\def\hatbolpi{\widehat{\boldsymbol{\pi}}}
 \global\long\def\hatbolrho{\widehat{\boldsymbol{\rho}}}
 \global\long\def\hatbolsigma{\widehat{\boldsymbol{\sigma}}}
 \global\long\def\hatboltau{\widehat{\boldsymbol{\tau}}}
 \global\long\def\hatbolphi{\widehat{\boldsymbol{\phi}}}
 \global\long\def\hatbolchi{\widehat{\boldsymbol{\chi}}}
 \global\long\def\hatbolpsi{\widehat{\boldsymbol{\psi}}}
 \global\long\def\hatbolomega{\widehat{\boldsymbol{\omega}}}
 \global\long\def\hatbolGamma{\widehat{\boldsymbol{\Gamma}}}
 \global\long\def\hatbolDelta{\widehat{\boldsymbol{\Delta}}}
 \global\long\def\hatbolTheta{\widehat{\boldsymbol{\Theta}}}
 \global\long\def\hatbolLambda{\widehat{\boldsymbol{\Lambda}}}
 \global\long\def\hatbolPi{\widehat{\boldsymbol{\Pi}}}
 \global\long\def\hatbolSigma{\widehat{\boldsymbol{\Sigma}}}
 \global\long\def\hatbolPhi{\widehat{\boldsymbol{\Phi}}}
 \global\long\def\hatbolPsi{\widehat{\boldsymbol{\Psi}}}
 \global\long\def\hatbolOmega{\widehat{\boldsymbol{\Omega}}}

\global\long\def\tilbolalpha{\widetilde{\boldsymbol{\alpha}}}
 \global\long\def\tilbolbeta{\widetilde{\boldsymbol{\beta}}}
 \global\long\def\tilbolgamma{\widetilde{\boldsymbol{\gamma}}}
 \global\long\def\tilboldelta{\widetilde{\boldsymbol{\delta}}}
 \global\long\def\tilbolepsilon{\widetilde{\boldsymbol{\epsilon}}}
 \global\long\def\tilbolzeta{\widetilde{\boldsymbol{\zeta}}}
 \global\long\def\tilboleta{\widetilde{\boldsymbol{\eta}}}
 \global\long\def\tilboltheta{\widetilde{\boldsymbol{\theta}}}
 \global\long\def\tilbolkappa{\widetilde{\boldsymbol{\kappa}}}
 \global\long\def\tilbollambda{\widetilde{\boldsymbol{\lambda}}}
 \global\long\def\tilbolmu{\widetilde{\boldsymbol{\mu}}}
 \global\long\def\tilbolnu{\widetilde{\boldsymbol{\nu}}}
 \global\long\def\tilbolxi{\widetilde{\boldsymbol{\xi}}}
 \global\long\def\tilbolpi{\widetilde{\boldsymbol{\pi}}}
 \global\long\def\tilbolrho{\widetilde{\boldsymbol{\rho}}}
 \global\long\def\tilbolsigma{\widetilde{\boldsymbol{\sigma}}}
 \global\long\def\tilboltau{\widetilde{\boldsymbol{\tau}}}
 \global\long\def\tilbolphi{\widetilde{\boldsymbol{\phi}}}
 \global\long\def\tilbolchi{\widetilde{\boldsymbol{\chi}}}
 \global\long\def\tilbolpsi{\widetilde{\boldsymbol{\psi}}}
 \global\long\def\tilbolomega{\widetilde{\boldsymbol{\omega}}}
 \global\long\def\tilbolGamma{\widetilde{\boldsymbol{\Gamma}}}
 \global\long\def\tilbolDelta{\widetilde{\boldsymbol{\Delta}}}
 \global\long\def\tilbolTheta{\widetilde{\boldsymbol{\Theta}}}
 \global\long\def\tilbolLambda{\widetilde{\boldsymbol{\Lambda}}}
 \global\long\def\tilbolPi{\widetilde{\boldsymbol{\Pi}}}
 \global\long\def\tilbolSigma{\widetilde{\boldsymbol{\Sigma}}}
 \global\long\def\tilbolPhi{\widetilde{\boldsymbol{\Phi}}}
 \global\long\def\tilbolPsi{\widetilde{\boldsymbol{\Psi}}}
 \global\long\def\tilbolOmega{\widetilde{\boldsymbol{\Omega}}}

\global\long\def\barbolmu{\overline{\bolmu}}
 \global\long\def\barmbX{\overline{\mbX}}

\global\long\def\mbbR{\mathbb{R}}
 \global\long\def\mbbS{\mathbb{S}}
 \global\long\def\mbbX{\mathbb{X}}
 \global\long\def\mbbY{\mathbb{Y}}
 \global\long\def\mbbZ{\mathbb{Z}}
 \global\long\def\mbbU{\mathbb{U}}

\global\long\def\calA{\mathcal{A}}
 \global\long\def\calB{\mathcal{B}}
 \global\long\def\calC{\mathcal{C}}
 \global\long\def\calD{\mathcal{D}}
 \global\long\def\calE{\mathcal{E}}
 \global\long\def\calF{\mathcal{F}}
 \global\long\def\calG{\mathcal{G}}
 \global\long\def\calH{\mathcal{H}}
 \global\long\def\calI{\mathcal{I}}
 \global\long\def\calJ{\mathcal{J}}
 \global\long\def\calK{\mathcal{K}}
 \global\long\def\calL{\mathcal{L}}
 \global\long\def\calM{\mathcal{M}}
 \global\long\def\calN{\mathcal{N}}
 \global\long\def\calO{\mathcal{O}}
 \global\long\def\calP{\mathcal{P}}
 \global\long\def\calQ{\mathcal{Q}}
 \global\long\def\calR{\mathcal{R}}
 \global\long\def\calS{\mathcal{S}}
 \global\long\def\calT{\mathcal{T}}
 \global\long\def\calU{\mathcal{U}}
 \global\long\def\calV{\mathcal{V}}
 \global\long\def\calW{\mathcal{W}}

\global\long\def\mbell{\boldsymbol{\ell}}
 \global\long\def\bolell{\boldsymbol{\ell}}
 \global\long\def\mbzero{\mathbf{0}}

\global\long\def\bolPhio{\boldsymbol{\Phi}_{0}}
 \global\long\def\bolOmegao{\boldsymbol{\Omega}_{0}}

\global\long\def\bolSigmaX{\bolSigma_{\mbX}}
 \global\long\def\bolSigmaY{\bolSigma_{\mbY}}
 \global\long\def\bolSigmaXY{\boldsymbol{\Sigma}_{\mbX\mbY}}
 \global\long\def\mbSX{\mathbf{S}_{\mbX}}
 \global\long\def\mbSY{\mathbf{S}_{\mbY}}
 \global\long\def\mbSXY{\mathbf{S}_{\mbX\mbY}}
 \global\long\def\mbSYX{\mathbf{S}_{\mbY\mbX}}
 \global\long\def\mbRYX{\mathbf{S}_{\mbY|\mbX}}
 \global\long\def\mbRXY{\mathbf{S}_{\mbX|\mbY}}
 \global\long\def\mbSc{\mbS_{\mbC}}
 \global\long\def\mbSd{\mbS_{\mbD}}

\global\long\def\sumn{\sum_{i=1}^{n}}

\global\long\def\E{\mathrm{E}}
 \global\long\def\F{\mathrm{F}}
 \global\long\def\J{\mathrm{J}}
 \global\long\def\H{\mathrm{H}}
 \global\long\def\G{\mathrm{G}}
 \global\long\def\Cov{\mathrm{cov}}
 \global\long\def\Corr{\mathrm{corr}}
 \global\long\def\Var{\mathrm{var}}
 \global\long\def\dimension{\mathrm{dim}}
 \global\long\def\spn{\mathrm{span}}
 \global\long\def\vech{\mathrm{vech}}
 \global\long\def\vecc{\mathrm{vec}}
 \global\long\def\Prob{\mathrm{Pr}}
 \global\long\def\Env{\mathrm{env}}
 \global\long\def\tr{\mathrm{tr}}
 \global\long\def\dg{\mathrm{diag}}
 \global\long\def\asyVar{\mathrm{avar}}
 \global\long\def\MSE{\mathrm{MSE}}
 \global\long\def\OLS{\mathrm{OLS}}

\global\long\def\sigerr{\sigma_{e}^{2}}
 \global\long\def\hatsigerr{\widehat{\sigma}_{e}^{2}}
 \global\long\def\bolSigmaf{\bolSigma_{\mbf}}
 \global\long\def\tilssigma{\widetilde{\sigma}^{2}}
 \global\long\def\hatssigma{\widehat{\sigma}^{2}}
 \global\long\def\ssigma{\sigma^{2}}
 \global\long\def\PLS{\mathrm{PLS}}
 \global\long\def\hatlambda{\widehat{\lambda}}
 \global\long\def\hatpi{\widehat{\pi}}

 \global\long\def\CRE{\mathcal{R}_{\bolSigma_{Y}}(\calB)}

\global\long\def\CS{\calS_{Y\mid\mbX}}
  \global\long\def\hatsigma{\widehat{\sigma}}
 \global\long\def\hatdelta{\widehat{\delta}}
 \global\long\def\hatb{\widehat{b}}
 \global\long\def\tilb{\widetilde{b}}

\newtheorem{lemma}{Lemma}
\newtheorem{proposition}{Proposition}
\newtheorem{theorem}{Theorem}
\newtheorem{definition}{Definition}
\newtheorem{example}{Example}

\newcommand{\beqn}{\begin{equation*}}
\newcommand{\eeqn}{\end{equation*}}

\newcommand{\bea}{\begin{eqnarray}}
\newcommand{\eea}{\end{eqnarray}}

\newcommand{\bean}{\begin{eqnarray*}}
\newcommand{\eean}{\end{eqnarray*}}

\newcommand{\beq}{\begin{equation}}
\newcommand{\eeq}{\end{equation}}

\newcommand{\texthl}[1]{{ {\color{blue} #1}}}

\title{{Covariate-Adjusted Tensor Classification in High-Dimensions}\thanks{\small The authors are grateful to the Editor, Associate Editor and two referees for insightful comments that led to significant improvements of our paper. The authors would like to thank Dr.~Lexin Li for sharing the ADHD and ASD data sets; and thank Drs.~Qun Li and Dan Schonfeld for sharing their code for methods CMDA and DGTDA. Research for this article was supported in part by grant CCF-1617691 and DMS-1613154 from the U.S.~National Science Foundation.  }}
\author{
{Yuqing Pan}\thanks{{\small {Yuqing Pan (yuqing.pan@stat.fsu.edu) is Ph.D.~student, Department of Statistics, Florida State University, Tallahassee, FL, 32306.
}}}
,\ {Qing Mai}\thanks{{\small {Qing Mai (mai@stat.fsu.edu) is Assistant Professor, Department of Statistics, Florida State University, Tallahassee, FL, 32306. }}}
,\ {and Xin Zhang}\thanks{{\small{Xin Zhang (henry@stat.fsu.edu) is Assistant Professor, Department of Statistics, Florida State University, Tallahassee,~FL, 32306.}}
}
}
\date{}
\maketitle
\begin{abstract}
In contemporary scientific research, it is of great interest to predict a categorical response based on a high-dimensional tensor (i.e. multi-dimensional array) and additional covariates. This mixture of different types of data leads to challenges in statistical analysis. Motivated by applications in science and engineering, we propose a comprehensive and interpretable discriminant analysis model, called CATCH model (in short for Covariate-Adjusted Tensor Classification in High-dimensions), which efficiently integrates the covariates and the tensor to predict the categorical outcome. The CATCH model jointly models the relationships among the covariates, the tensor predictor, and the categorical response. More importantly, it preserves and utilizes the structures of the data for maximum interpretability and optimal prediction. To tackle the new computational and statistical challenges arising from the intimidating tensor dimensions, we propose a penalized approach to select a subset of tensor predictor entries that has direct discriminative effect after adjusting for covariates. We further develop an efficient algorithm that takes advantage of the tensor structure. Theoretical results confirm that our method achieves variable selection consistency and optimal classification error, even when the tensor dimension is much larger than the sample size. The superior performance of our method over existing methods is demonstrated in extensive simulated and real data examples.


\vspace*{.2in}

\end{abstract}
\newpage{}
%
\section{Introduction}

Many contemporary scientific and engineering studies collect data from different categories of subjects in the form of multiple-dimensional array, a.k.a.~tensor, accompanied by additional covariates. 
For example, an important application area of our proposed method is neuro-imaging analysis, where researchers often want to identify and understand neurological and neuro-developmental disorders from a discriminant analysis model built on tensor images, such as anatomical magnetic resonance imaging (MRI), positron emission tomography (PET), functional magnetic resonance imaging (fMRI), and electroencephalography (EEG), plus a few additional clinical covariates such as medical measurements and psychological and cognitive scores. 
This type of data also frequently arise in computational biology, personalized recommendation, and image recognition analysis, among others.

The increasing popularity of such data brings many new challenges to statisticians. First, it is generally unclear how to integrate the information from both the tensor-variate predictor and the vector of covariates to achieve the best possible classification. The tensors and the covariates may affect each other, and how to model their dependence on each other and define their effects on the response remains an open question.  Secondly, the tensor predictor is often high-dimensional. For example, in our neuroimaging data applications, we use the structural magnetic resonance imaging (MRI) to study the attention deficit hyperactivity disorder (ADHD) and autism spectrum disorder (ASD). Each MRI is a three-way tensor with dimension $30\times36\times30$ (ADHD) or $91\times109\times91$ (ASD), which is more than $30,000$ or $900,000$ entries for one subject. Moreover, rapid advancements in neuroimaging technology enable researchers to obtain tensor images with higher and higher resolutions and hence higher dimensions. This calls for new high-dimensional algorithms and methods that scale well with the increasing tensor dimensions.
Thirdly, it is non-trivial to extend the vector-based high-dimensional statistical properties and theoretical results to high-dimensional higher-order tensor predictors.

In this article, we study the discriminant analysis with a high-dimensional tensor predictor $\mbX\in\mbbR^{p_1\times\cdots\times p_M}$, $M\geq2$, a low-dimensional covariates vector $\mbU\in\mbbR^q$, and a class label $Y\in\{1,\dots,K\}$ for $K\geq2$ categories.
For such problems, we propose a unified framework called the CATCH model that jointly utilizes information from the entire tensor and the covariates, while the intrinsic tensor-on-covariates relationship is accounted for through a regression model. We carefully investigate the direct and indirect effects of $\mbU$ on $Y$. The direct effect of $\mbU$ helps separate classes. Meanwhile, correctly adjusting for the indirect effect of $\mbU$ through an intrinsic tensor regression model of $\mbX$ on $\mbU$ may substantially improve estimation, variable selection and the prediction of $Y$. We further identify the direct effect of $\mbX$ on discriminating $Y$, after adjusting for $\mbU$. With a limited sample size, it is necessary to perform some type of dimension reduction on the adjusted $\mbX$. Conceptually, our reduction of tensor covariates is similar to that of the partial dimension reduction methods in regression \citep[e.g.]{partialSDR2, partialSDR3}, where we want to reduce the dimension of the predictor without losing any information in classification after adjusting for the covariates effects.

While numerous high-dimensional classification methods have been developed, they may still not scale well with tensor data, because most of them are designed for vector data. As previously mentioned, in many neuroimaging studies, simply vectorizing the tensor image results in a vector of length in the order of $10^5\sim10^7$. Moreover, many high-dimensional sparse classification methods \citep[e.g.]{Cai2011,ROAD,Xu2015} require computing the sample covariance of this vector, which has over $10^{10}\sim10^{14}$ entries. This is apparently very computationally demanding. An intuitive remedy for this issue is to perform marginal screening \citep[e.g]{Pan2016} on the tensor predictor. Although marginal screening is computationally efficient, it is well-known that marginally important predictors may not be jointly important, and vice versa. More importantly, both the vectorization approach and the marginal screening approach ignore the tensor structure and hence may lose important structural information. As we show in this paper, discarding the tensor structure deprives an opportunity of reducing the number of parameters. Because of these issues, it is important to develop a method that models the entire tensor without sacrificing its tensor structure to preserve interpretability. While there has been an enormous body of literature on sparse linear discriminant analysis (LDA) for high-dimensional (vector) predictor \citep[e.g.]{Cai2011, Shao2011, Clemmensen, witten, ROAD, DSDA,Xu2015,MSDA}, with the tensor structure and the additional covariates to be adjusted for, we are facing a much more complicated high-dimensional problem, which requires a new statistical model, more efficient and scalable algorithms and more involved theoretical studies. 

Our proposal is related to but fundamentally different from recent developments in tensor regression and tensor decomposition. Although classification is one of the most common statistical tasks, it receives relatively less attention than regression in the research of tensor data. Many researchers have studied matrix- and tensor-variate regression \citep[e.g.]{zhou2013tensor, zhou2014regularized, zhao2014structured,  hoff2015multilinear, raskutti2015convex, sun2016provable, wang2016generalized, LiZhang2015, zhang2016tensor, lock2017tensor}. But most of these methods do not directly apply to classification or incorporating the covariates. {We propose a general framework for joint modeling and multi-class classification with both tensor predictor and vector covariates. Under this framework, we develop a new method that achieves optimal classification and consistent variable selection in high-dimensional tensor coefficients.} Moreover, many existing statistical methods on tensor data rely on multi-linear tensor decomposition \citep[e.g.]{KoldaBader09Tensor, chi2012tensors, liu2017characterizing, zhang2017guaranteed} that assumes low-rank structures of the tensor. Our approach does not require any low-rank approximation of the tensor predictor. Instead, we directly identify and eliminate the unimportant tensor discriminative coefficients in our model and thus achieve variable selection and parsimonious modeling. {Our sparsity pursuit on tensor discriminative coefficients provides a good alternative to the popular low-rank and sparse-low-rank techniques. On one hand, the rank determination of low-rank tensor decomposition is a very challenging problem that usually brings more tuning parameters, while underestimated ranks would lead to bias and loss on some subtle tensor information. Without adopting any low-rank approximation/assumption, our approach of variable selection in tensor coefficients is more direct and flexible. On the other hand, the  penalization approach proposed in this paper can be easily adjusted by specifying different penalty terms on different regions of the tensor to incorporate prior information such as smoothness, regions of interests, and regions of gray or white matters in brain images.}

In the literature, most of the matrix/tensor discriminant analysis methods have their roots in Fisher's discriminant analysis. For a matrix or tensor predictor, various approaches \citep{Zhong2015,GTDA,DATER,STDA,CMDA,tensornetwork,Zeng2015} are proposed to find linear projections on each mode of the tensor to have the maximum between-class separation with respect to within-class variability. Although these methods were developed in a similar context as our proposal, there are significant distinctions. First, existing methods typically do not consider how to incorporate information from these additional covariates. Second, our proposal is based on a probabilistic model instead of being motivated by maximizing between-class variability. Consequently, our method, to the best of our knowledge, is the first in the tensor discrimination analysis literature to provide strong theoretical guarantees of (i) recovering the Bayes' rule, (ii) consistently selecting important tensor discriminantive entries, (iii) algorithm convergence, with ultra-high dimensional tensors. In our numerical studies, we have also confirmed superb performances of our method in terms of classification accuracy, variable selection, and computational time. Logistic regression is another popular approach for tensor classification. For example, \citep{zhou2013tensor} adopted tensor low-rank structures in a generalized linear model, \citet{Wimalawarne} proposed to add various tensor norms as penalties to the logistic loss. These methods only handles binary classification, while CATCH is naturally applicable to multiclass problems. In addition, unlike our theoretical studies, the theoretical results in \citet{Wimalawarne} only concern the logistic loss but not classification error or variable selection.

The contributions of this article are multi-fold. First of all, it addresses the important question of how to jointly model and explain the relationships among a mixed type  of data: categorical response, continuous multivariate covariates and high-dimensional tensor predictor. Our CATCH model offers a useful solution by systematically and simultaneously studying the tensor-on-covariate regression, and the covariate-on-response, tensor-on-response classifications. Secondly, while existing high-dimensional classification methods concentrate on a vector predictor, our work extends the scope of applications to high-dimensional tensor. To achieve such an important extension, we have developed new computational and theoretical techniques. Thirdly, our proposal greatly advances the recent development of tensor data analysis. While existing approaches largely rely on tensor regression and especially tensor low-rank decomposition, we focus on discriminant analysis and classification. Our method provides an alternative way of tensor dimension reduction by introducing group sparsity directly based on the Bayes' rule and hence achieves optimal classification. 

The rest of this paper is organized as follows. We review some tensor notations in Section~\ref{review}. In Section~\ref{sec:catch_model}, we introduce the CATCH model and define the direct and indirect effects in the model. In particular, the potential gain in classification from adjusting for covariates are discussed in Section~\ref{Bayes' rule}. In Section~\ref{sec:estimation}, we discuss how to estimate the Bayes' rule under the CATCH model for classification. 
In Section~\ref{sec:algorithm}, we develop an efficient algorithm that actively takes advantage of the tensor structure so that we can conduct the computation with minimal storage. 
Section~\ref{sec:theory} contains theoretical studies of both non-asymptotic and asymptotic properties of the proposed method in ultra-high dimensional settings. Extensive simulations in Section~\ref{sec:sim} and two real data applications in Section~\ref{sec:realdata} confirm the advantages of our method over existing methods. Finally, Section~\ref{sec:discussion} contains a short discussion and the Supplementary Materials contain additional numerical studies, along with proofs and other technical details.

\subsection{Review of some tensor notations}\label{review}

We first introduce some standard tensor notations and operations that are used frequently in this manuscript and are standard in the tensor literature \citep[for example]{KoldaBader09Tensor}.

For positive integers $M\geq2$, $p_1,\ldots,p_M$, a multidimensional array $\mbA\in \mbbR^{p_{1}\times\cdots\times p_{M}}$ is referred to as an \emph{$M$-way} or \emph{$M$-th order tensor}. The vectorization of a tensor $\mbA$, $\vecc(\mbA)$, is a $(\prod_{m}p_m\times1)$ column vector, with $A_{i_1\cdots i_M}$ being its $j$-th element, $j=1+\sum_{k=1}^M(i_k-1)\prod_{k'=1}^{k-1}p_{k'}$. The \emph{mode-$k$ matricization}, $\bA_{(k)}$, is a $(p_k\times \prod_{m\ne k}p_m)$ matrix, with $A_{i_1\cdots i_M}$ being its $(i_k,j)$-th element, $j=1+\sum_{k'=k}(i_{k'}-1)\prod_{l<k',l\ne k}p_{l}$. 
If we fix every index of the tensor but one, then we have a \emph{fiber}. For example, $A_{i_1\cdots i_{k-1}I_ki_{k+1}\cdots i_M}$, $I_k=1,\dots p_k$, form a $(p_k\times1)$ vector called the mode-$k$ fiber of $\mbA$. 
The \emph{mode-$k$ product} of a tensor $\mbA$ and a matrix $\bolalpha\in\mbbR^{d\times p_{k}}$, denoted by $\mbA\times_{k}\bolalpha$, is a $M$-way tensor of dimension $p_{1}\times\cdots\times p_{k-1}\times d\times p_{k+1}\times\cdots\times p_M$, with each element being the product of a mode-$k$ fiber of $\mbA$ and a row vector of $\bolalpha$.
The \emph{mode-$k$ vector product} of a tensor $\mbA$ and a vector $\mbc\in\mbbR^{p_k}$, denoted by $\mbA\bar\times_{k}\mbc$ is a $(M-1)$-way tensor of dimension $p_1\times\cdots\times p_{k-1}\times p_{k+1}\times\cdots \times p_M$, with each element being the inner product of a mode-$k$ fiber of $\bA$ and $\mbc$. The \emph{Tucker decomposition} of a tensor is defined as $\mbA = \mbC\times_{1}\mbG_1\times_{2}\cdots\times_{M}\mbG_M$, where $\mbC\in\mbbR^{d_{1}\times\cdots\times d_{M}}$ is the \emph{core tensor}, and $\mbG_k\in\mbbR^{p_{k}\times d_{k}}$, $k=1,\dots,M$, are the \emph{factor matrices}. We write the Tucker decomposition as $\llbracket\mbC;\mbG_1,\dots,\mbG_m\rrbracket$ in short. In particular, we frequently use the fact that $\vecc(\llbracket\mbC;\mbG_1,\dots,\mbG_M\rrbracket)=\left(\mbG_M\otimes\cdots\otimes\mbG_1\right)\vecc(\mbC)=\left(\bigotimes_{m=M}^{1}\mbG_m\right)\vecc(\mbC)$, where $\otimes$ denotes Kronecker product and $\bigotimes_{m=M}^{1}\mbG_m$ is short for $\mbG_M\otimes\cdots\otimes\mbG_1$.

We introduce the tensor normal (TN) distribution as a generalization of the matrix normal distribution \citep{gupta1999matrix}. For a  tensor random variable $\bZ\in\mathbb{R}^{p_1\times\cdots\times p_M}$, it is called a standard tensor normal random variable if all elements of $\mbZ$ independently follow the (univariate) standard normal distribution. If $\bX=\bmu+\llbracket \bZ; \bSigma_1^{1/2},\ldots,\bSigma_M^{1/2}\rrbracket$, we say $\mbX$ follows a tensor normal distribution $\bX\sim TN(\bmu, \bSigma_1,\ldots,\bSigma_{M})$, where $\bSigma_j>0$ imposes the dependence structure on the $j$-th mode. Hence, $\vecc(\mbX)=\vecc(\bolmu) + \bolSigma^{1/2}\vecc(\mbZ)$, where $\bolSigma=\bigotimes_{m=M}^{1}\bolSigma_m$.

\section{The CATCH Model}\label{sec:catch_model}

\subsection{The model assumptions}\label{CATCH}

We propose the CATCH (covariates-adjusted tensor classification in high dimensions) model for a random triplet $\{Y,\bU,\bX\}$, where $Y\in\{1,\dots,K\}$ is the class label for $K\geq2$ classes, $\bU\in\mathbb{R}^{q}$ is a vector of covariates that needs to be adjusted for, and $\mbX\in\mathbb{R}^{p_1\times \cdots\times p_M}$ is a $M$-th order tensor-variate predictors, $M\geq2$. Throughout this paper, we assume that $\Prob(Y=k)=\pi_{k}>0$ where $\sum_{k=1}^{K}\pi_{k}=1$. Our goal is to build a classifier that accurately predicts $Y$ based on integrated information from $\mbU$ and $\mbX$. 
To this end, we propose the CATCH model:
\begin{eqnarray}\label{CATCH_eq1}
\mbU\mid(Y=k) & \sim & N(\bolphi_{k},\bolPsi),\\
\mbX\mid(\mbU=\mbu,Y=k) & \sim & TN(\bolmu_{k}+\bolalpha\bar{\times}_{(M+1)}\mbu,\bolSigma_{1},\dots,\bolSigma_{M}),\label{CATCH_eq2}
\end{eqnarray}
where $\bolphi_{k}\in\mbbR^{q}$, $\bolPsi\in\mathbb{R}^{q\times q}$, $\bolPsi>0$ is symmetric, $\bolalpha\in\mbbR^{p_{1}\times\cdots\times p_{M}\times q}$, $\bolmu_k\in\mathbb{R}^{p_1\times \cdots p_M}$, and $\bSigma_{m}\in\mathbb{R}^{p_m\times p_m}$, $\bSigma_m>0$ is symmetric, $m=1,\ldots, M$. It is obvious that all the parameters have natural interpretation. In \eqref{CATCH_eq1}, we assume that $\{Y,\mbU\}$ follows the classical LDA model, where $\bolphi_k$ is the mean of $\mbU$ within class $k$ and $\bolPsi$ is the common within class covariance of $\mbU$. Similarly, in \eqref{CATCH_eq2}, we assume a common within class covariance structure of $\mbX$ characterized by $\bSigma_{m}, m=1,\ldots, M$, that does not depend on $Y$ after adjusting for the covariates $\mbU$. The tensor coefficient $\bolalpha$ characterizes the linear dependence of the tensor predictor $\mbX$ on the covariates $\mbU$, and $\bolmu_k$ is the covariate-adjusted within-class mean of $\mbX$ in class $k$.

{Although our CATCH model is based on discriminant analysis models, which may seem stringent, many existing results in the literature support their applications in practice. For example, \citet{michie1994machine,hand2006classifier} reported that LDA is competitive on many benchmark datasets, while \citet{Cai2011, Shao2011, Clemmensen, witten, ROAD, DSDA,Xu2015} demonstrated the competitive classification performance of sparse LDA methods on high-dimensional datasets. These encouraging results lead us to consider the models in \eqref{CATCH_eq1}--\eqref{CATCH_eq2} for tensor classification. Our real data analysis in Section~\ref{sec:realdata} also confirms that the classifier based on the CATCH model achieves accurate results in practice comparing to many well-known classifiers. Hence, we expect our classifier to be widely applicable, while model assumptions such as normality are imposed to provide intuition. Meanwhile, from the statistical perspective, it would still of great interest to develop classifiers under weaker model assumptions. See Section~\ref{sec:discussion} for some discussion along this line. We leave this topic for future research.}

An important special case of the CATCH model applies to the situation when we only have $\{\mbX,Y\}$, but not the covariate $\mbU$. Then \eqref{CATCH_eq1}--\eqref{CATCH_eq2} reduce to
\beq\label{TDA}
\mbX\mid(Y=k)\sim TN(\bolmu_{k},\bolSigma_{1},\dots,\bolSigma_{m}),
\eeq
which implies that the tensor predictor $\mbX$ follows the tensor normal distribution with different means but common covariance structure. We refer to the model in \eqref{TDA} as the tensor discriminant analysis (TDA) model. It is a natural extension of LDA to incorporate tensor structure, and is different from modeling $\{Y,\vecc(\mbX)\}$ using the classical LDA. By utilizing the tensor structure, we greatly reduce the number of free parameters. In the LDA model on predictor $\vecc(\mbX)$ of dimension $\prod_{m=1}^Mp_m\times1$, the covariance matrix has $\prod_{m=1}^M p_m^2$ elements. In contrast, the covariance structure in \eqref{TDA} takes advantage of the tensor structure and only has $\sum_{m=1}^M p_m^2$ elements. In Section~\ref{sec:algorithm}, we show that this structure leads to convenience in computation. 

When both covariates and tensor are present, the CATCH model not only characterizes how the covariates $\mbU$ and the tensor variable $\mbX$ simultaneously distinguish the classes $Y$, but also model the regression relationship of $\mbX$ on $\mbU$ within each class.
To gain more insights, within each class $k$, we can write \eqref{CATCH_eq2} as 
\begin{equation}\label{TensorResponseRegression}
\mbX=\bolmu_k+\bolalpha\bar\times_{(M+1)}\mbU+\mbE,\quad \mbE\sim TN(0,\bolSigma_{1},\dots,\bolSigma_{M}),
\end{equation}
where $\mbE$ is an unobservable tensor normal error independent of $\mbU$.
Equation \eqref{TensorResponseRegression} coincides with the tensor response regression (TRR) model proposed by \citet{LiZhang2015}. The tensor parameter $\bolmu_k$ is the adjusted mean of $\mbX$ in class $k$ after removing the effect of covariates $\mbU$ on $\mbX$. Estimation and inference of $\bolalpha$ and $\bolmu_j-\bolmu_k$, $j\neq k$, are of great interest in neuroimaging analysis and applications, where $\bolalpha$ describes the effect of covariates and $\bolmu_j-\bolmu_k$ compares tensor images across classes after adjusting for covariates. Although the focus of this paper is not studying the interrelationship between $\mbX$ and $\mbU$, accounting for this intrinsic regression relation \eqref{TensorResponseRegression} often brings substantial gain in predicting $Y$, in estimating discriminative parameters, and even in variable selection. We explain this phenomenon in the following section, right after we define the direct and indirect effects.

\begin{figure}[t!]
\begin{center}
  \begin{tikzpicture}
    [scale=1.8,auto=right]
    \node[vertex][label=below:Covariates] (v1) at (1,4)  {$\mathbf{U}$};
    \node[vertex][label=Tensor] (v2) at (4,6.2)  {$\mathbf{X}$};
    \node[vertex][label=below:Class Label] (v3) at (7,4)  {$Y$};

    \draw[-{Stealth[scale=1.8]}] (v1) to node [above,sloped]{{\Large $\bolalpha$}}(v2);
        \draw[-{Stealth[scale=1.8]}] (v1) to node [below,sloped]{TRR model \eqref{TensorResponseRegression}}(v2);
    \draw[-{Stealth[scale=1.8]}] (v2) to node [above,sloped] {{\Large $\mathbf{B}_2,\cdots,\mathbf{B}_K$}} (v3);
        \draw[-{Stealth[scale=1.8]}] (v2) to node [below,sloped] {TDA model \eqref{TDA} } (v3);
    \draw[-{Stealth[scale=1.8]}] (v1) to node {{\Large $\bolgamma_2,\cdots,\bolgamma_K$}} (v3);
        \draw[-{Stealth[scale=1.8]}] (v1) to node [above]{LDA model \eqref{CATCH_eq1} } (v3);
  \end{tikzpicture}
  \end{center}
  \caption{\label{fig:CATCH_effect} Graphical illustration of the direct and indirect effects in the CATCH model. The direct effect of $\mbU$ on $Y$ reflects the classical linear discriminant analysis (LDA;~\ref{CATCH_eq1}) model; the direct effect of $\mbX$ on $Y$ reflects the tensor discriminant analysis (TDA;~\ref{TDA}) model; the indirect effect of $\mbU$ on $Y$ through affect $\mbX$ resembles the tensor response regression (TRR;~\ref{TensorResponseRegression}). }
\end{figure}
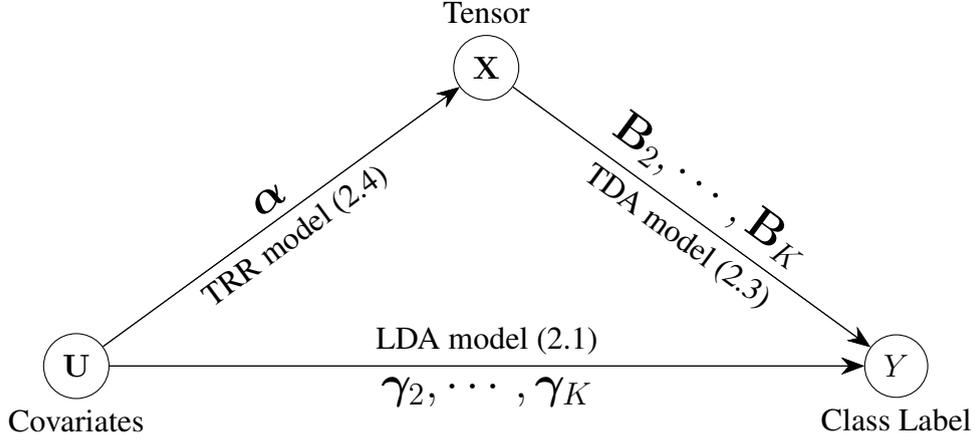

\subsection{The direct and indirect effects} \label{Bayes' rule}

Since our goal is to predict $Y$ based on $\{\mbU,\mbX\}$, we derive the ideal classifier -- the so called ``Bayes' rule'' -- under the CATCH model. Estimation of this classifier is discussed later in Section~\ref{sec:estimation}. 
Given $\mbX$ and $\mbU$, the Bayes' rule that achieves the lowest error rate possible is defined as (e.g. \citet{FHT01}),
\beq\label{BR-general}
\widehat Y=\arg\max_{k=1,\ldots,K} \Pr(Y=k\mid \mbX=\mbx,\mbU=\mbu)=\arg\max_{k=1,\ldots,K} \pi_kf_k(\mbx,\mbu),
\eeq
where $\pi_k=\Pr(Y=k)$ and $f_k(\mbx,\mbu)$ is the joint probability density function of $\mbX$ and $\mbU$ conditional on $Y=k$. We have the following results under the CATCH model.

\begin{proposition}\label{BR}
The Bayes' rule of CATCH model (\ref{CATCH_eq1},~\ref{CATCH_eq2}) is
\begin{equation}
\widehat{Y}=\arg\max_{k=1,\dots K}\left\{ a_k+\bolgamma_{k}\T\mbU+\langle\mbB_{k},\mbX-\bolalpha\bar{\times}_{(M+1)}\bU\rangle\right\},\label{CATCH_Bayes}
\end{equation}
where $\bgamma_k=\bolPsi^{-1}(\bolphi_k-\bolphi_1)$, $\mbB_k=\llbracket \bmu_k-\bmu_1; \bSigma_1^{-1},\ldots,\bSigma_M^{-1}\rrbracket$,  and $a_k=\log({\pi_k}/{\pi_1})-\frac{1}{2}\bolgamma_k\T(\bolphi_k+\bolphi_1)-\langle \mbB_k,\frac{1}{2}(\bolmu_k+\bolmu_1)\rangle$ is a scalar that does not involve $\mbX$ or $\mbU$.
\end{proposition}

The parameters $\{\bolgamma_k,\bolalpha,\mbB_k\}$ can be viewed as the direct and indirect effects of $\mbU$ and the direct effect of $\mbX$ after adjusting for $\mbU$, respectively.  First, the discriminative coefficient vector $\bolgamma_k\in\mbbR^q$ is the direct effect of $\mbU$ on classification and coincides with the usual LDA discriminative directions in \eqref{CATCH_eq1}. Second, the tensor regression coefficient $\bolalpha\in\mbbR^{p_1\times\cdots\times p_M\times q}$ is the indirect effect of $\mbU$ on $Y$. It characterizes how $\mbU$ affects $Y$ through its relationship with $\mbX$. Finally, the discriminative coefficient tensor $\mbB_k\in\mbbR^{p_1\times\cdots\times p_M}$ is the direct effect of $\mbX$ after adjusting for $\mbU$. By Proposition~\ref{BR}, in absence of the covariates $\mbU$, the Bayes' rule of the TDA model \eqref{TDA} is $\widehat{Y}=\arg\max_{k=1,\dots K}\left\{ \log(\pi_{k}/\pi_{1})+\langle\mbB_{k},\mbX-\frac{1}{2}(\bmu_k+\bmu_1)\rangle\right\}$, where $\mbB_k$ is defined in Proposition~\ref{BR}. A graphical illustration of the direct and indirect effects is in Figure~\ref{fig:CATCH_effect}.

When the covariates $\mbU$ have different means $\bolphi_k$ in each class, they directly contribute to the separation of the classes along the discriminative directions $\bolgamma_k=\bolPsi^{-1}(\bolphi_k-\bolphi_1)$. 
However, somewhat surprisingly, even when the covariates have no direct effect on separating classes, i.e. $\bolphi_1=\cdots=\bolphi_K$, the inclusion of $\mbU$ can still bring substantial gain in classification.

When $\bolphi_1=\cdots=\bolphi_K$, $\mbU$ still has indirect effect on classification through affecting $\mbX$ by $\bolalpha\overline{\times}_{M+1}\mbU$. This can be seen from comparing the Bayes' error based on the Bayes' rule in \eqref{CATCH_Bayes}. Define $R(\mbU,\mbX)$ as the lowest classification error rate possible if we build the classifier based on $\mbX$ and $\mbU$, and similarly, $R(\mbX)$ as that based only on $\mbX$. The explicit expressions of $R(\mbX)$ and $R(\mbU,\mbX)$ are given in the Supplementary Materials. The following toy example demonstrates that, ignoring the indirect effect of the covariates greatly inflates the classification error and changes the importance of predictors.

\begin{figure}[t!]
\begin{center}
\includegraphics[scale=0.8, trim=0.5cm 9cm 0cm 9cm]{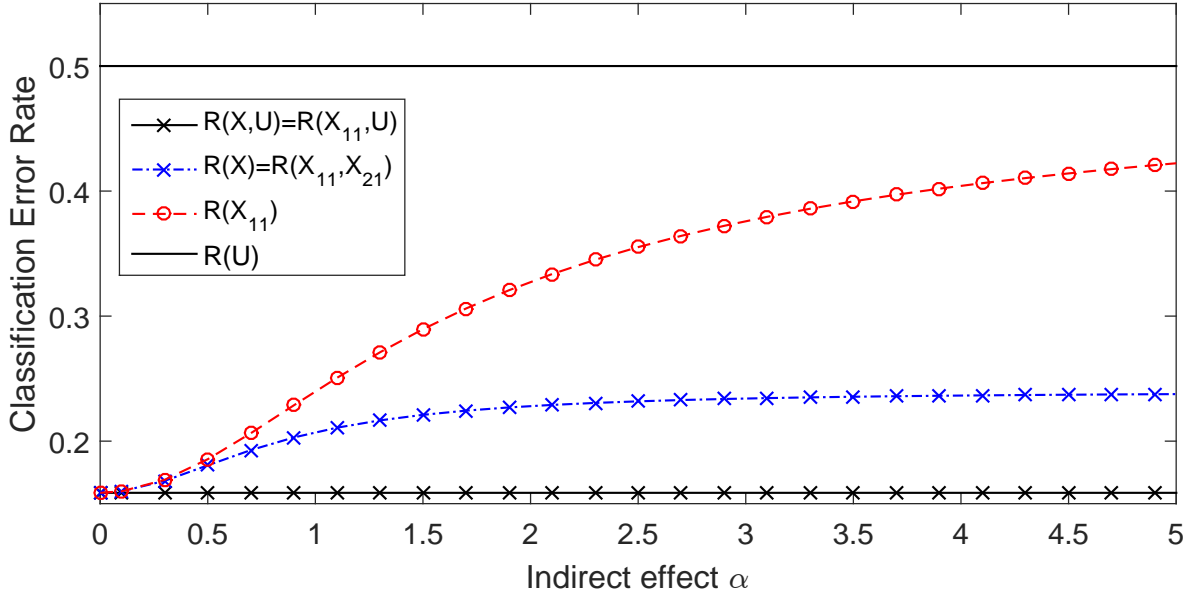}
\end{center}
\caption{\label{fig:illustration} Graphical illustration of the best possible classification error rates in Example 1. }
\end{figure}

\begin{example}\label{example1}
Consider a binary classification example $Y=1$ or $2$ with equal class probability $\pi_1=\pi_2=0.5$, where the covariates $U\mid(Y=1)\sim U\mid(Y=2)\sim N(0,1)$ has no direct effect on classifying $Y$. The tensor $\mbX$ is a $2\times2$ matrix, and $\E(\mbX\mid U=u,Y=k)=\bolmu_k+\bolalpha\cdot u$, where $\bolmu_1=\begin{pmatrix}
0 & 0 \\
0 & 0
\end{pmatrix}$, $\bolmu_2=\begin{pmatrix}
2 & 0 \\
0 & 0
\end{pmatrix}$ and $\bolalpha=\begin{pmatrix}
\alpha & 0 \\
\alpha & 0
\end{pmatrix}$, and covariance $\bolSigma_1=\bolSigma_2=\mbI_2$. 
Under this model, the discriminative coefficient matrix $\mbB_2=\bolSigma_1^{-1}(\bolmu_2-\bolmu_1)\bolSigma_2^{-1}$ is zero everywhere except for its first element, indicating that only $X_{11}$ has direct effect on classification. If we ignore $U$, then $\mbX\mid (Y=k)$ is no longer a matrix normal random variable but $\vecc(\mbX)\mid (Y=k)$ is multivariate normal with mean $\vecc(\bolmu_k)$ since $\E(U\mid Y)=0$, and covariance $\bolSigma=\bolSigma_2\otimes\bolSigma_1+\vecc(\bolalpha)\vecc\T(\bolalpha)$. By straightforward calculation (see Supplementary Materials), we have $R(U)=0.5$, $R(\mbX,U)=R(X_{11},U)=1-\Phi(1)=0.1587$, where $\Phi(\cdot)$ is the cumulative distribution function $N(0,1)$. If we ignore $\mbU$, $R(\mbX)=R(X_{11},X_{21})=1-\Phi(\sqrt{2+\alpha^{2}}/\sqrt{2+2\alpha^{2}})<R(X_{11})=1-\Phi(1/\sqrt{1+\alpha^2})$. As the number $\alpha\rightarrow\infty$ in the indirect effect $\bolalpha$, the error rates $R(X_{11})\rightarrow0.5$ and $R(\mbX)\rightarrow1-\Phi(1/\sqrt{2})=0.2398$.
\end{example}

We further plotted the error rates using different predictors versus $\alpha$ in Figure~\ref{fig:illustration}.
The Bayes' error is always $R(\mbX,U)=R(X_{11},U)=0.1587$, indicating that the magnitude of indirect effect $\bolalpha$ does not affect the classification error if we adjust for $U$ correctly. When we fail to adjust for $U$, $R(X_{11})$ increases drastically with $\alpha$ and eventually converges to $0.5$; and $R(\mbX)=R(X_{11},X_{21})$ increases quickly with $\alpha$ and eventually converges to around $0.2398$, which is much larger than the Bayes' error of $0.1587$ with $U$.  In this example, to achieve the best classification error, we only need one element of the tensor predictor, $X_{11}$ if we have adjusted for $U$, but we need two elements $X_{11}$ and $X_{21}$ if not. 
Example~\ref{example1} hence exhibits the potential impact of the covariates on variable selection of $\mbX$: the best possible classifier based on $\mbX$ may have a different sparsity pattern, depending on whether we adjust for covariates correctly. 

\subsection{The tensor discriminative set and the multi-class group sparsity}\label{DDS}

To estimate the Bayes' rule in Proposition~\ref{BR}, we need to estimate $\bolalpha$ and $\{\pi_k,\bolphi_k,\bolgamma_k,\bolmu_k,\mbB_k\}_{k=1}^K$. By definition, we have $\bolgamma_1=0,\mbB_1=0$, so we only need to estimate $\bolgamma_k,\mbB_k$ for $k=2,\ldots,K$. In this paper we focus on low-dimensional covariates and high-dimensional tensor predictor, although it is possibly straightforward to generalize our proposal to incorporate high-dimensional $\mbU$. Henceforth, we assume that the sample size $n$ satisfies $q<n\ll \prod_{m=1}^M p_m$.

Since $\mbU$ is low-dimensional, the estimation of parameters $\bolphi_k$, $\bolgamma_k$ and $\bolPsi$ is relatively straightforward. On the other hand, although $\bolalpha$ is high-dimensional, it is connected to the tensor regression model and can be estimated easily under the $q<n$ scenario. However, the estimations of tensor coefficients $\mbB_k$, $k=2,\dots,K$, are more challenging since they are high-dimensional and depend on the covariance structures $\bolSigma_1,\dots,\bolSigma_M$. In practice we typically do not have a sufficient sample size to accurately estimate all the $(K-1)\cdot\prod_{m=1}^M p_m$ coefficients in $\mbB_2,\dots,\mbB_K$ without additional assumptions. It is well-received that the sparsity assumption is crucial in high-dimensional classifications \citep[e.g.]{BL04, fan2008high}.

From Proposition~\ref{BR} and our discussions, an entry of the tensor predictor $X_{j_1\cdots j_M}$ has an effect on the final classification (after adjusted for the covariates) if and only if $b_{k,j_1\cdots j_M}\ne 0$ for some $k$, where $b_{k,j_1\cdots j_M}$ is the $(j_1\cdots j_M)$-th entry of $\mbB_k$. Hence, we introduce our notion of tensor discriminative set in the CATCH model that leads to the sparsity assumption of the model. Define the discriminative set $\mathcal{D}$ and its complement set $\mathcal{D}^c$ as follows,
\begin{eqnarray}
{\cal D}&=&\{(j_1,\ldots,j_M): b_{k,j_1\cdots j_M}\ne 0 \mbox{ for some $k$}\},\label{Dset}\\
{\cal D}^c&=&\{(j_1,\ldots,j_M): b_{1,j_1\cdots j_M}=\cdots=b_{K,j_1\cdots j_M}=0\}.\label{DCset}
\end{eqnarray}

The sparsity assumption then requires that the cardinality (the number of nonzero entries) of ${\cal D}$, denoted as $d=\vert\mathcal{D}\vert$, is much smaller than the dimension $\prod_{m=1}^M p_m$, so that most of the predictors belong to the complement set ${\cal D}^c$. 

{Examination of ${\cal D}^c$ reveals that the coefficients in $\mbB_2,\ldots\mbB_K$ have a group sparsity structure across classes, because for any $(j_1,\ldots,j_M)$, the coefficients $(b_{2,j_1\cdots J_M},\ldots,b_{K,j_1\cdots j_M})$ are all coefficients for one voxel $X_{j_1\cdots j_M}$; they are the effects of $X_{j_1\cdots j_M}$ in separating different pairs of classes. When $X_{j_1\cdots j_M}$ is not important, i.e., does not have effect in separating any pair of classes, all its coefficients have to be 0. Consequently, we have the group sparsity structure across classes (rather than across voxels). For vector data, \citet{hastie2015statistical} considered a similar group sparsity assumption across classes in multinomial regression, which shares some spirit with our assumption. We remark here, though, that the group structure is present only when $K>2$.  When $K=2$, we only need one set of coefficients $\mbB_2$ to separate two classes and $\vert\calD\vert$ becomes the number of nonzeros in $\mbB_2$. It follows that \eqref{Dset} \& \eqref{DCset} reduce to ${\cal D}=\{(j_1,\ldots,j_M):b_{2,j_1\cdots j_M}\ne 0\}$ and ${\cal D}^C=\{(j_1,\ldots,j_M):b_{2,j_1\cdots j_M}= 0\}$, which resembles the more familiar form of sparsity, such as that in regression problems \citep{Tibshirani1996}.}


\section{Estimation Procedure}\label{sec:estimation}
In this section, we assume that we have obtained i.i.d.~samples $\{Y^{i},\mbU^{i},\mbX^{i}\}_{i=1}^n$ and discuss how to build an accurate classifier based on the data. With a little abuse of notation, we set $\bY\in\mathbb{R}^n$ as a vector that contains all the observed class labels, $\mbU\in\mathbb{R}^{n\times q}$ as a matrix that contains all the observed covariates and $\mbX\in \mathbb{R}^{p_1\times \cdots\times p_M\times n}$ as a $(M+1)$-way tensor data. AS we have discussed  in Section~\ref{DDS}, the sparsity assumption is only imposed on $\mbB_k$ but not on other parameters $\bolalpha, \{\pi_k,\bolphi_k,\bgamma_k,\bolmu_k\}_{k=1}^K$. 
We hence separately discuss the un-penalized estimations of $\bolalpha, \{\pi_k,\bolphi_k,\bgamma_k,\bolmu_k\}_{k=1}^K$ in Section~\ref{subsec:unpenalized} and $\{\bolSigma_m\}_{m=1}^M$ in Section \ref{subsec:estSig} and the penalized estimation of $\mbB_2,\dots,\mbB_K$ in Section~\ref{subsec:penalized}.

\subsection{Estimation of $\{\pi_k,\bolphi_k,\bgamma_k,\bolmu_k\}_{k=1}^K$ and $\bolalpha$}\label{subsec:unpenalized}

We let $\overline{\mbU}_k$ be the sample mean of $\mbU$ within Class $k$, and $\barmbX_k$ be the sample mean of $\mbX$ within Class $k$. We estimate $\{\pi_k,\bolphi_k,\bgamma_k\}$ straightforwardly using the following sample estimators, which are maximum likelihood estimators (MLE) under the CATCH model (\ref{CATCH_eq1},~\ref{CATCH_eq2}),
\begin{equation}\label{MLE_gamma}
\widehat\pi_k=\dfrac{1}{n}\sum_{i=1}^n \mathrm{1}(Y^i=k),\quad
\hatbolphi_k=\overline{\mbU}_k,\quad
\widehat\bolgamma_k=\widehat\bolPsi^{-1}(\hatbolphi_k-\hatbolphi_1),\quad
k=1,\dots,K,
\end{equation}
where $\widehat\bolPsi=\dfrac{1}{n}\sum_{k=1}^K\sum_{y^i=k}(\mbU^i-\hatbolphi_k)(\mbU^i-\hatbolphi_k)\T$. 

Meanwhile, the MLE for $\bolalpha$ can be most succinctly expressed using tensor products and the group-wise centered data: for the observations within class $k$, i.e. $Y^i=k$, let $\tilmbX^{i}=\mbX^{i}-\barmbX_{k}$ and $\widetilde{\mbU}^{i}=\mbU^{i}-\overline{\mbU}_k$ and define $\widetilde{\mbX}\in\mbbR^{p_{1}\times\cdots\times p_{M}\times n}$ and
$\widetilde{\mbU}\in\mbbR^{q\times n}$ to be the tensor and the matrix that consist of $\tilmbX^{i}$ and $\widetilde{\mbU}^{i}$, respectively.
\begin{lemma}\label{MLE}
Under the CATCH model (\ref{CATCH_eq1},~\ref{CATCH_eq2}), the maximum likelihood estimator of $\bolalpha$ is
\beq\label{alpha.MLE}
\widehat{\bolalpha}=\widetilde{\mbX}\times_{(M+1)}\{(\widetilde{\mbU}\widetilde{\mbU}\T)^{-1}\widetilde{\mbU}\}.
\eeq
\end{lemma}

Note that we assume $q\ll n$. Hence, $(\widetilde{\mbU}\widetilde{\mbU}\T)\in\mbbR^{q\times q}$ is invertible and \eqref{alpha.MLE} is a legitimate estimate for CATCH model. To gain more intuition of this estimate, note that the CATCH model in \eqref{CATCH_eq2} implies that, for each $X_{j_1\cdots j_M}$, within class $k$, we have
\beq
X_{j_1\cdots j_M}-\mu_{k,j_1\cdots j_M}=\balpha_{j_1\cdots j_M}\T\mbU+\epsilon_{j_1\cdots j_M}
\eeq
where the vector $\balpha_{j_1\cdots j_M}$ is a mode-$(M+1)$ fiber of $\bolalpha$, and $\epsilon_{j_1\cdots j_M}$ is a normal random variable with mean zero and is independent of $\mbU$. Hence, within each class, each entry in the tensor depends on the covariates through a linear regression model. Meanwhile, \eqref{alpha.MLE} implies that
\beq\label{alpha.MLE1}
\widehat{\bolalpha}_{j_1\cdots j_M}=\left\{ \sum_{k=1}^{K}\sum_{Y^i=k}(\mbU^{i}-\overline{\mbU}_{k})(\mbU^{i}-\overline{\mbU}_{k})\T\right\} ^{-1}\left\{ \sum_{k=1}^{K}\sum_{Y^i=k}(\mbU^{i}-\overline{\mbU}_{k})(X^{i}_{j_1\cdots j_M}-\overline{X}_{k,j_1\cdots j_M})\right\}
\eeq
The estimate in \eqref{alpha.MLE1} closely resembles the ordinary least squares estimate in linear regression, except that both $\mbU^i$ and $\mbX^i$ are centered within their individual classes. This distinction comes from the fact that, our CATCH model implies the tensor response regression models \eqref{TensorResponseRegression} within each class $k$. Therefore, we need to adjust $\mbU$ and $\mbX$ by their within class mean. 

The connection of $\hatbolalpha_{j_1,\ldots,j_M}$ with the least squares estimator suggests an easy extension for estimating $\bolalpha_{j_1,\ldots,j_M}$ when the covariates are also high-dimensional with $q\gg n$. For example, in disease diagnostic studies based on both the brain images and genetics data, we can replace the least squares estimator with the penalized least squares estimator on $(\widetilde{\mbU}^i,\widetilde{X}_{j_1\cdots j_M} )$. 


To estimate the intercept $\bolmu_k$ based on the tensor response model \eqref{TensorResponseRegression}, we have
\beq\label{muk.hat}
\hatbolmu_k=\overline{\mbX}_k-\hatbolalpha\bar{\times}_{(M+1)}\overline{\mbU}_k,\quad k=1,\dots,K,
\eeq
where $\hatbolalpha$ is obtained in \eqref{alpha.MLE}.

\subsection{Estimation of $\{\bolSigma_m\}_{m=1}^{M}$\label{subsec:estSig}}

To estimate $\bolSigma_m, m=1,\ldots,M$, we derive the following Lemma~\ref{lem:cov}. A similar result for matrix normal distribution has been presented in \citet{gupta1999matrix}.

\begin{lemma}\label{lem:cov}
If $\mbW\sim TN(0,\bolOmega_1,\ldots,\bolOmega_m)$, then 
\beq
\E\{\mbW_{(j)}\mbW_{(j)}\T\}=\bolOmega_j\cdot\prod_{l\ne j} \tr(\bolOmega_{l}),
\eeq
where $\mbW_{(j)}$ is the mode-$j$ matricization of $\mbW$.
\end{lemma}
A direct implication of Lemma~\ref{lem:cov} is that we can obtain an unbiased estimator for $\bolSigma_m$ up to a scale change, based on the fitted residuals from \eqref{TensorResponseRegression}. { From \eqref{TensorResponseRegression} and \eqref{muk.hat}, we have the fitted residuals for all $i,k$, such that $Y^i=k$,
$$
\widehat{\mbE}^{i}=\mbX^{i}-\hatbolmu_k+\hatbolalpha\bar\times_{(M+1)}\mbU^{i}=(\mbX^{i}-\overline{\mbX}_k)-\hatbolalpha\bar\times_{(M+1)}(\mbU^{i}-\overline{\mbU}_k),
$$
where the second equality using ``centered'' variables facilitates implementation as we no longer need to use $\hatbolmu_k$'s in our implementation. } We define 
$\widetilde{\mbS}_j=(n\prod_{l\neq j}^{M}p_l)^{-1}\sum_{i=1}^{n}\widehat{\mbE}^{i}_{(j)}(\widehat{\mbE}^{i}_{(j)})\T$. 
Then according to Lemma~\ref{lem:cov}, $\E(\widetilde{\mbS}_j)=c \bolSigma_j$ for some scalar $c$. To properly scale $\widetilde{S}_j$, we have
\bea\label{sigma.hat}
\hatbolSigma_{j}=\widetilde s_{j,11}^{-1}\widetilde{\mbS}_j \mbox{ for $j=1,\ldots,M-1$}; \hatbolSigma_{M}=\dfrac{\widehat{\Var}(X_{1 \cdots 1})}{\prod_{j=1}^M \widetilde s_{j,11}}\widetilde{\mbS}_M.
\eea

{It is easy to see that $\widehat{\bolSigma}_j$ is always positive semi-definite. But we have the further result concerning the positivity of $\widehat{\bolSigma}_j$ in the following lemma. 
\begin{lemma}\label{lem.pd}
If 
\beq\label{dim.pd}
(n-K)\prod_{m\ne j}p_{m}>p_j,
\eeq
then $\widehat\bSigma_j$ is positive definite with probability 1.
\end{lemma}
It follows that, if \eqref{dim.pd} holds, our penalized optimization introduced later in \eqref{TLDA-formula} is strictly convex with a probability of 1. Later we will see that this result helps with the convergence analysis for our algorithm. It is also worth noting that, the condition in \eqref{dim.pd} is very mild when the dimensions of each mode $p_m, m=1,\ldots,M$ are roughly comparable. For example, if $p_1=\ldots=p_M$, the condition in \eqref{dim.pd} is true as long as $n-K>1$. Meanwhile, if \eqref{dim.pd} does not hold, we could always perturb $\widehat{\bolSigma}_j$ as follows:
\beq\label{sigma.plus.i}
\widehat{\bolSigma}_j'=\widehat{\bolSigma}_j+\gamma\bI_{p_j}
\eeq
where $\gamma>0$ is a small constant. A similar estimator has been considered in the vector case \citep{ledoit2004well} to guarantee positivity of the covariance estimator. Plugging in the estimator in \eqref{sigma.plus.i} results in a strictly convex optimization problem.
}

We would like to remark here that many other proposals exist for estimating $\bolSigma_j$ \citep{Dutilleul1999, ManceurDutilleul2013, werner2008Kroncov}. While other estimators can be directly used as a plug-in to our CATCH optimization \eqref{TLDA-formula}, they are generally more computationally demanding than our estimator. 
Because the parameters $\bSigma_m,m=1,\ldots, M$ are nuisance to the Bayes' rule (c.f. Proposition~\ref{BR}), the estimation of them is an intermediate step to constructing estimates of $\mbB_k$. Therefore, we use the estimator in \eqref{sigma.hat} for easy computation. Also, we will show in Section~5 that they will eventually lead to consistent estimate of $\mbB_k$ in high dimensions.

\subsection{Penalized estimation of $\{\mbB_k\}_{k=2}^{K}$}\label{subsec:penalized}

{By Proposition~\ref{BR}, $\mbB_k=\llbracket\bmu_k-\bmu_1;\bSigma_1^{-1},\ldots,\bSigma_M^{-1} \rrbracket\in\mathbb{R}^{p_1\times\cdots\times p_M}, k=2,\ldots, K$. To facilitate sparse estimation, we rewrite $\{\mbB_k\}_{k=2}^{K}$ as the solution to an optimization problem as follows.}
{
\begin{lemma}\label{B:opt}
For $\mbC_k\in\mathbb{R}^{p_1\times\cdots\times p_M}, k=2,\ldots, K$, define the objective function 
\beq
\calL(\mbC_2,\ldots,\mbC_K)=\sum_{k=2}^K\{\langle\mbC_k,\llbracket\mbC_k;\bolSigma_{1},\dots,\bolSigma_{M}\rrbracket\rangle-2\langle\mbC_k,\bolmu_k-\bolmu_{1}\rangle\}.
\eeq
Then $(\mbB_2,\cdots,\mbB_K)=\arg\min_{\mbC_2,\ldots,\mbC_K}\calL(\mbC_2,\ldots,\mbC_K)$.
\end{lemma}
}

{Lemma~\ref{B:opt} implies that the un-penalized estimators $\widetilde\mbB_k\equiv\llbracket\hatbolmu_k-\hatbolmu_1;\hatbolSigma_1^{-1},\dots,\hatbolSigma^{-1}_M\rrbracket$, $k=2,\dots,K$, must be the solution to the following quadratic optimization problem, 
\begin{equation}\label{B.pop}
(\tilmbB_2,\ldots,\tilmbB_K)=\arg\min_{\mbB_2,\dots,\mbB_K}\sum_{k=2}^K\{\langle\mbB_k,\llbracket\mbB_k;\hatbolSigma_{1},\dots,\hatbolSigma_{M}\rrbracket\rangle-2\langle\mbB_k,\hatbolmu_k-\hatbolmu_{1}\rangle\},
\end{equation}
where the sample estimators $\hatbolmu_k,k=1,\ldots,K,\hatbolSigma_m,m=1,\ldots,M$ are obtained in previous sections. Finally, our CATCH estimators $(\hatmbB_2,\ldots,\hatmbB_K)$ are defined as the minimizers of the following penalized estimation,}
\begin{equation}\label{TLDA-formula}
\min_{\mbB_2,\ldots,\mbB_K}\left[\sum_{k=2}^K\left(\langle\mbB_k,\llbracket\mbB_k;\hatbolSigma_{1},\dots,\hatbolSigma_{M}\rrbracket\rangle-2\langle\mbB_k,\hatbolmu_{k}-\hatbolmu_{1}\rangle\right)
+\lambda\sum_{j_{1}\dots j_{M}}\sqrt{\sum_{k=2}^{K}b_{k,j_{1}\cdots j_{M}}^2}\right],
\end{equation}
where $\lambda>0$ is a tuning parameter. Compared with the original quadratic optimization in the population, \eqref{B.pop}, we have added the group LASSO penalty \citep{yuan2006model} to the sample optimization because of the group sparsity structure across groups in $\mbB_2,\dots,\mbB_K$ as discussed in Section~\ref{DDS}. The penalty reduces to the LASSO penalty \citep{Tibshirani1996} when $K=2$.  Large values of $\lambda$ encourage group sparsity among $\mbB_2,\dots,\mbB_K$ at matching coordinates in the discriminative set $\mathcal{D}$, e.g. $\widehat b_{2,j_1\cdots j_M}=\cdots=\widehat b_{K,j_1\cdots j_M}=0$. 
With an appropriate $\lambda$, we will have a consistent estimate of $\mathcal{D}$, i.e. $\widehat{\mathcal{D}}=\mathcal{D}$, with probability 1 as established later in Theorem~\ref{consistency}.

\section{Algorithm and Its Convergence}\label{sec:algorithm}

All the estimates except for $\hatmbB_k$ in Section~\ref{sec:estimation} can be implemented straightforwardly. Here, we propose an algorithm for estimating $\mbB_k$ based on \eqref{TLDA-formula} that scales well with high dimensions.

For convenience, define $\bolbeta_k=\mathrm{vec}(\mbB_k)$, $\bolnu_k=\vecc(\bolmu_k)$ and $\hatbolnu_k=\vecc{(\hatbolmu_k)}$. Rewrite our problem (\ref{TLDA-formula}) with $\bolbeta_k$ as our parameters as
\begin{equation}\label{sparse obj vecBk}
(\hatbolbeta_2,\ldots,\hatbolbeta_K)=\arg\min_{\bolbeta_2,\ldots,\bolbeta_K}\bigg[\sum_{k=2}^K\{\bolbeta_k\T(\hatbolSigma_{M}\otimes\cdots\otimes\hatbolSigma_{1})\bolbeta_k
-2(\hatbolnu_k-\hatbolnu_{1})^T\bolbeta_k+\lambda\sum_{j=1}^{p}\Vert\bolbeta_{\boldsymbol{\cdot} j}\Vert\}\bigg],
\end{equation}
where $p=\prod_{m=1}^M p_m$, $\bolbeta=(\bolbeta_2,\dots,\bolbeta_K)\T\in\mbbR^{(K-1)\times p}$, $\bolbeta_{\boldsymbol{\cdot} j}\in\mbbR^{K-1}$ denotes the $j$-th column vector of $\bolbeta$, and $\Vert\bolbeta_{\boldsymbol{\cdot} j}\Vert=(\sum_{k=2}^{K}\beta_{kj}^{2})^{\frac{1}{2}}$. After obtaining $\hatbolbeta_k$, the CATCH estimator $\hatmbB_k$ is obtained by mapping $\hatbolbeta_k$ back to the original tensor structure.

At first glance, \eqref{sparse obj vecBk} is a penalized quadratic problem. In particular, if we ignore the Kronecker product structure and simply let $\widehat\bSigma=\hatbolSigma_{M}\otimes\cdots\otimes\hatbolSigma_{1}$, then $\eqref{sparse obj vecBk}$ reduces to
\begin{equation}\label{vecBk no Kronecker}
(\hatbolbeta_2,\ldots,\hatbolbeta_K)=\arg\min_{\bolbeta_2,\ldots,\bolbeta_K}\bigg[\sum_{k=2}^K\{\bolbeta_k\T\hatbolSigma\bolbeta_k-2(\hatbolnu_k-\hatbolnu_{1})^T\bolbeta_k+\lambda\sum_{j=1}^{p}\Vert\bolbeta_{\boldsymbol{\cdot} j}\Vert\}\bigg],
\end{equation}
which resembles the objective function of multiclass sparse discriminant analysis in \citet{MSDA} and can be solved by the algorithm therein when $\hatbolSigma$ is not huge. However, for high-dimensional tensors, the dimension of $\hatbolSigma$ is $\prod_{m=1}^M p_m \times \prod_{m=1}^M p_m$. Even the storage of such a huge matrix can be challenging, let alone further operations on it. Therefore, we propose a new algorithm that takes advantage of the Kronecker product structure of $\hatbolSigma=\bigotimes_{m=M}^{1}\hatbolSigma_m$.

Define the operator $(x)_+=x$ if $x\ge 0$ and $(x)_+=0$ if $x<0$. The $\mathrm{mod}$ operator is defined by modulo operation, however, we let $a$ $\mathrm{mod}$ $b=b$ if the reminder of a modulo $b$ is 0. We also define two sequences of numbers for each $j$:
\bea
&&s_{M+1}=j, s_m=s_{m+1}\ \mathrm{mod}\ \prod_{i=1}^{m-1} p_i, \mbox{ for $m=M,\ldots,2$}\\
&&j_1=s_2\ \mathrm{mod}\ p_1, j_m=\lceil\frac{s_{m+1}}{\prod_{i=1}^{m-1}p_i}\rceil, \mbox{ for $m=2,\ldots, M$}\label{jm}
\eea
We need the two sequences $\{s_m\},\{j_m\}$ for technical reasons. See Lemma D.1 in the Supplementary Materials for more details. Then our algorithm is based on the following results. 

\begin{lemma}\label{lem:CD}
For $m=1,\ldots,M$, define $\hatbolSigma_{m, \boldsymbol{\cdot}j}$ as the $j$'th column of $\hatbolSigma_{m}$.
For each $j=1,\ldots,p$, the solution to $\bolbeta_{\cdot j}$ to (\ref{sparse obj vecBk}) given ${\bolbeta_{\boldsymbol{\cdot} j'},j'\neq j}$, 
is the same as
\begin{equation}\label{CD}
\arg\min_{\bolbeta_{\cdot j}}\sum_{k=2}^K\frac{1}{2}(\beta_{kj}-\widetilde\beta_{kj})^2+\frac{\lambda}{\hatsigma_{jj}}\Vert\bolbeta_{\boldsymbol{\cdot} j }\Vert,
\end{equation}
where 
\beq\label{beta.tilde}
\widetilde{\beta}_{kj}=\dfrac{(\widehat{\nu}_{kj}-\widehat{\nu}_{k1})-\llbracket \mbB_k^{j};\hatbolSigma\T_{1, \boldsymbol{\cdot}j_1},\cdots,\hatbolSigma\T_{M, \boldsymbol{\cdot}j_M}\rrbracket}{
\prod_{m=1}^M\widehat\sigma_{m,j_mj_m}},
\eeq
{with $\widehat{\sigma}_{m,j_mj_m}$ being the $(j_m,j_m)$-th element of $\widehat{\bSigma}_m$,} and $\mbB_k^j\in\mbbR^{p_1\times\cdots\times p_M}$ is a tensor such that $\vecc{(\mbB_k^j)}_{j'}$ equals $\beta_{kj'}$ for $j'\ne j$ and 0 otherwise. The indices $j_1,\ldots,j_M$ are defined in \eqref{jm}. Finally, the solution for \eqref{CD} is
\begin{equation}\label{hatbolbeta.CD}
\hatbolbeta_{\boldsymbol{\cdot} j}=\widetilde{\bolbeta}_{\boldsymbol{\cdot} j}\left(1-\frac{\lambda}{\Vert\widetilde{\bolbeta}_{\boldsymbol{\cdot} j}\Vert}\right)_+.
\end{equation}
\end{lemma}

\begin{algorithm}[t!]
\begin{enumerate}
\item Input $\hatbolSigma_m, m=1,\ldots,M$ and $\hatbolmu_k, k=1,\ldots,K$. Initialize $\hatmbB_k=0$ for all $k=2,\ldots, K$.

\item For steps $w=1,2,\ldots$, do the following until convergence:

for each element $j=1,\ldots,p$,
\begin{enumerate}
\item Update $\mbB_k^j$ based on current $\hatbolbeta_{\cdot j'}^{(w-1)}$ for all $j'\neq j$ and compute 
\beq\label{update}
\widehat\beta_{kj}^{(w)}\longleftarrow\dfrac{(\widehat{\nu}_{kj}-\widehat{\nu}_{k1})-\llbracket \mbB_k^{j};\hatbolSigma\T_{1, \boldsymbol{\cdot}j_1},\cdots,\hatbolSigma\T_{M, \boldsymbol{\cdot}j_M}\rrbracket}{
\prod_{m=1}^M\widehat\sigma_{m,j_mj_m}}
\eeq
\item Compute $\widehat\beta_{kj}^{(w)}\longleftarrow\widetilde \beta_{kj}^{(w-1)}\left(1-\dfrac{\lambda}{\sqrt{\sum_{k=2}^{K}(\widetilde\beta_{kj}^{(w-1)})^{2}}}\right)_{+}$
for $k=2,\ldots,K$
\end{enumerate}
\item Output $\hatmbB_{k}$, where $\vecc{(\hatmbB_k)}=\hatbolbeta_{k}$ at convergence.
\end{enumerate}
\caption{\label{alg:sparseLDA} Algorithm for CATCH}
\end{algorithm}


By Lemma~\ref{lem:CD}, we only need to iterate over $j$ to solve for $\widetilde{\bolbeta}_{\cdot j}$ and $\widehat{\bolbeta}_{\cdot j}$ to obtain $\widehat\mbB_k,k=2,\ldots,K$. Hence, we propose Algorithm~\ref{alg:sparseLDA} to solve for $\widehat\mbB_k$. To implement our algorithm, in each iteration we only need certain columns of $\hatbolSigma_1,\ldots,\hatbolSigma_M$ based on Lemma~\ref{lem:CD}, because we have taken into account the Kronecker structure in \eqref{beta.tilde}. Hence, the space required to implement our algorithm is of the order $O(\sum_{m=1}p_m^2)$. In contrast, if we ignore the Kronecker structure and solve \eqref{vecBk no Kronecker}, we will have to compute $\hatbolSigma$ beforehand, which requires the space at the order of $O(\prod_{m=1}^Mp_m^2)$. By taking advantage of the Kronecker product structure, we gain considerable saving in space. Our algorithm scales much better to high-dimensional tensor data. 

{We also have the following result for per-iteration computational complexity.
\begin{lemma}\label{lem:computationalcost}
The computational cost for updating $\widehat\beta_{kj}^{(t)}$ is $O(MKd_tp)$, where $d_t$ is the number of nonzero coefficients at iteration $t$, $t=1,2,\dots$.
\end{lemma}
As for computational complexity, in each iteration the cost is $O(d_tp)$, where $d_t$ is the number of nonzero coefficients in the iteration.} 

{
Finally, we present the convergence result for our blockwise coordinate descent algorithm. For ease of presentation, we assume that \eqref{dim.pd} holds and hence our optimization problem is strictly convex with a probability of 1. If \eqref{dim.pd} does not hold, we can always replace the covariance estimates by those in \eqref{sigma.plus.i} to achieve similar convergence results.
\begin{theorem}\label{thm:convergence}
If $(n-K)\prod_{j\neq m}p_{j}\ge p_m$, with a probability of 1, our blockwise coordinate descent algorithm converges to the global minimizer of \eqref{sparse obj vecBk}.
\end{theorem}
Theorem~\ref{thm:convergence} shows that there is no gap between the output of our algorithm and the global minimizer of \eqref{sparse obj vecBk}. This fact facilitates the theoretical studies of CATCH, as will be presented in the next section.
}

\section{Theory}\label{sec:theory}

In this section, we study the statistical properties of CATCH.  Theorem~\ref{consistency} establishes the consistency of the direct effect estimation, i.e. $\hatmbB_k\rightarrow\mbB_k$, $k=1,\dots,K$, and the consistency of the discriminative set recover, i.e. $\widehat{\mathcal{D}}\rightarrow \mathcal{D}=\{(j_1,\ldots,j_M): b_{k,j_1\cdots j_M}\ne 0 \mbox{ for some $k$}\}$, where $\widehat{\mathcal{D}}$ is the estimated discriminative set based on our sparse estimator $\hatmbB_k$, $k=1,\dots,K$. 
Theorem~\ref{prediction} establishes the optimal prediction of our method: the classification error rate of CATCH converges to the Bayes' error rate.

We introduce some notations. For a matrix $\bV\in\mathbb{R}^{q_1\times q_2}$, $\Vert\bV\Vert_{\infty}=\max_{i}\sum_{j=1}^{q_2}|v_{ij}|$, $\Vert\bV\Vert_{1}=\max_{j}\sum_{i=1}^{q_1}|v_{ij}|$. For an $m$-way tensor $\bW\in \mathbb{R}^{q_1\times \cdots\times q_m}$, denote $\Vert\bW\Vert_{\max}=\max_{j_1,\cdots, j_m}|w_{j_1\cdots j_m}|$.
Throughout this section $C$ denotes a generic positive constant that could vary from line to line. We use $\bolSigma=\bigotimes_{m=M}^{1}\bolSigma_m$ to simplify the presentation of the theoretical results, although in practice we never directly use the estimate of $\bolSigma$ (c.f. Algorithm~\ref{alg:sparseLDA}). We also let $p=\prod_{m=1}^M p_m, p_{-m}=\prod_{j\ne m} p_j$, $d=|{\cal D}|$ and
\bea 
b_{\max}&=&\max_{k,j_1\cdots j_M}|b_{k,j_1\cdots j_M}|\\
b_{\min}&=&\min_{(k,j_1\cdots j_M):b_{k,j_1\cdots j_M}\ne 0}|b_{k,j_1\cdots j_M}|\\
\varphi&=&\max\{\Vert\bSigma_{{\cal D}^C,{\cal D}}\Vert_{\infty},\Vert\bSigma^{-1}_{{\cal D},{\cal D}}\Vert_{\infty}\}\\
\Delta&=&\max\{\Vert \left(\vecc(\bmu_1),\cdots,\vecc(\bmu_K)\right)\Vert_1,\Vert \left(\vecc(\bB_2),\cdots,\vecc(\bB_K)\right)\Vert_1\}
\eea

For simplicity, we make a few assumptions about the parameters, but all the assumptions in this paragraph can be relaxed, at the cost of more lengthy proofs.  The number of classes, $K$, the number of covariates, $q$, and the order of the tensor, $M$, are all assumed to be fixed. We assume that $\Vert\bSigma_j^{1/2}\Vert_{1}$, $\Vert \bolphi_k\Vert_{\max}$ and $\Vert\bmu_k\Vert_{\max}$ are bounded above uniformly with respect to $p$. We further assume that the diagonal elements of $\bSigma_j$, $j=1,\dots,M$, are all ones.

 The following technical conditions will be used in the theorems.
\begin{enumerate}
\item[(C1)] $\max_{j\in {\cal D^C}}\left\{\sum_{k=2}^K(\bolSigma_{j{\cal D}}\bolSigma_{{\cal D}{\cal D}}^{-1}\mbt_{k{\cal D}})^2\right\}^{1/2}=\kappa<1$, where $\mbt_{k\mathcal{D}}$ is defined as the sub-gradient of the group lasso penalty term in the objective function \eqref{sparse obj vecBk} with respect to $\bolbeta_{k\mathcal{D}}$.
\item[(C2)] There exists $c_1>0$ such that $\pi_k\ge\dfrac{c_1}{K}$ for $k=1,\ldots,K$.
\item[(C3)] The largest eigenvalues of $\bSigma_m,m=1,\ldots,M$, are uniformly bounded above by a constant $C_2>0$. 
\item[(C4)] $\min_{j,k}\{(\vecc(\bolmu_k-\bolmu_j))\T\bSigma^{-1}(\vecc(\bolmu_k-\bolmu_j))\}$ is bounded away from $0$.
\item[(C5)] $\left\{\dfrac{d^2(\log{d}+\sum_{m=1}^M\log{p_m})}{n}\right\}^{1/2}=o(b_{\min})$ {as $n\rightarrow \infty$}.
\end{enumerate}

{Condition (C1) is a technical assumption similar to the standard condition in group lasso penalized regression model \citep{bach2008consistency}. Condition (C2) implies that the classes are reasonably balanced and as the sample size increases, each class will have a reasonably large sample size. Condition (C3) mimics a popular assumption in high-dimensional data analysis. For example, in \citet{CL2011} where they considered sparse linear discriminant analysis, it was assumed that the largest eigenvalue of the covariance matrix is bounded above. Condition (C4) guarantees that the classes are well separated to allow for accurate prediction and error rate consistency. Condition (C5) imposes a constraint on the dimensions and the signal strength. If $b_{\min}=O(1)$ and $d=O(n^{\xi})$ for $0<\xi<1/2$, then we can allow $\log{p_m}=o(n^{1-2\xi})$. Hence, we can allow the dimension of each mode of the tensor dimension to grow at an exponential rate of the sample size. Meanwhile, we can also allow $b_{\min}$ to decay at a rate determined by the dimensionality.}

\begin{theorem}[Estimation and Variable Selection Consistency]\label{consistency}
Under Conditions (C1)--(C3), there exists a generic constant $\psi$ such that, if $0<\lambda<\min\{\dfrac{b_{\min}}{8\varphi},\psi(1-\kappa),1\}$, then with a probability greater than
$$
1-2K\sum_{m=1}^M p_{m}^2\exp(-\dfrac{Cnp_{-m}\lambda^2}{d^2})-CKq^2p\exp\{-\dfrac{Cn\lambda^2}{d^2q^2}\}-2Kp\exp(-Cn\dfrac{\lambda^2}{d^2}),
$$ 
we have that $\widehat{\cal D}={\cal D}$ and $\Vert \vecc(\widehat\bB_k)-\vecc(\bB_k)\Vert_{\infty}\le 4\varphi\lambda$.
If we further assume Condition (C5), and that $\{\dfrac{d^2(\log{d}+\sum_{m=1}^M\log{p_m})}{n}\}^{1/2}\ll \lambda\ll b_{\min}$, $\lambda\rightarrow 0$, then we have the following statements with a probability tending to 1, 
$$\widehat{\cal D}={\cal D},\quad \Vert\vecc(\widehat\bB_k)-\vecc(\bB_k)\Vert_{\infty}\rightarrow 0,\quad k=1,\dots,K.$$ 
\end{theorem}

Theorem~\ref{consistency} implies that, under Conditions (C1)--(C3), (C5) and (C6), we can correctly identify the important features and accurately estimate the discriminant effects with a probability tending to 1. This supports the application of our proposed method in high-dimensional data.

We also remark here that the proofs in Theorem~\ref{consistency} are much more involved than those in the sparse linear discriminant analysis literature \citep{CL2011,ROAD,DSDA,MSDA} for two reasons. First, we have tensor normal data and we estimate the covariance by a Kronecker product of marginal sample covariances. Consequently, the existing results for sample covariance do not apply here. A relevant paper is \citet{Zhou2014} where the author presented large deviation results for matrix normal distribution. But in the current manuscript we show large deviation results for tensor normal distribution, without relying on any of the results in \citet{Zhou2014}, which can be of independent interest. Secondly, we have two layers of hierarchical structures. When we estimate the parameters for tensor data, we have to resort to the pseudo data $\bX-\hatbolalpha\bar{\times}_{(M+1)}\mbU$, where $\hatbolalpha$ introduces additional noise. There is no such issue in sparse LDA. Fortunately, in our careful theoretical studies, we observe that the estimation error introduced by $\hatbolalpha$ is usually of a higher order than the estimation error in estimation based on $\bX-\bolalpha\bar{\times}_{(M+1)}\mbU$. This assures that although we have to estimate $\bolalpha$, it has very little effect on our final estimation. Such results also support our unpenalized estimation procedure for $\bolalpha$.

In what follows, we further present results concerning the classification error rates. Since we observe that $\hatbolalpha$ is generally a good surrogate for $\bolalpha$, we consider the simplified case where we only have tensor data but not covariates. In the rest of this section, we assume the model in \eqref{TDA}. For a new observation $\{\bX^{\text{new}},Y^{\text{new}}\}$ not involved in fitting the classifier. Define the classification error rate of our CATCH estimator and that of the Bayes rule as follows:
\begin{eqnarray}
R_n 
&=&
\Pr\left(\widehat{Y}(\mbX^{\text{new}}\mid\hatmbB_k,\widehat{\pi}_k,\hatbolmu_k)\neq Y^{\text{new}}\right),\\
R 
&=&
\Pr\left(\widehat{Y}(\mbX^{\text{new}}\mid\mbB_k,{\pi}_k,\bolmu_k)\neq Y^{\text{new}}\right).
\end{eqnarray}
Clearly, we hope $R_n$ to be as close to $R$ as possible. Indeed, in the following theorem we show that $R_n$ converges to $R$ with an overwhelming probability.

\begin{theorem}[Optimal Prediction and Bayes' Rule Consistency]\label{prediction}
Under 
Conditions (C1)--(C4), there exists a generic constant $\psi_1>0$ such that, if $0<\lambda<\min\{\dfrac{b_{\min}}{8\varphi},\psi_1(1-\kappa),1\}$, then with a probability greater than 
\beq
1-2K\sum_{m=1}p_{m}^2\exp(-C\dfrac{np_{-m}\lambda^2}{d^2})-CKq^2p\exp\{-\dfrac{Cn\lambda^2}{d^2q^2}\}-2Kp\exp(-Cn\dfrac{\lambda^2}{d^2}),
\eeq
we have
$$
|R_n-R|\le C\lambda^{1/3}.
$$
If we further assume Condition (C5), $\dfrac{d^2(\log{d}+\sum_{m=1}^M\log{p_m})}{n}\ll\lambda\ll b_{\min}$ and $\lambda\rightarrow 0$, then with a probability tending to 1, 
\beq
R_n\rightarrow R.
\eeq
\end{theorem}

According to Theorem~\ref{prediction}, CATCH can asymptotically achieve the best classification accuracy. Therefore, CATCH is a powerful prediction tool as well.

\section{Simulations}\label{sec:sim}

In this section, we present numerical results to demonstrate the performance of the CATCH estimator. In Section~\ref{subsec:without covs} we present simulation results in problems with matrix and three-way tensor predictors but no covariates so that we can compare different treatments to the tensor data. In Section~\ref{covariates} we present simulation results in problems where both tensor predictors and covariates are present to investigate the importance of adjusting for the covariates. All simulations in Section~\ref{covariates} are binary classification problems, $K=2$; and simulations in Section~\ref{subsec:without covs} are multi-class problems with $K=3$ or $4$.

We include various popular and state-of-the-art classification methods as competitors. From machine learning and high-dimensional statistics literature, we include $\ell_1$-penalized Fisher's discriminant analysis \citep[$\ell_1$-FDA;][]{witten}, sparse optimal scoring \citep[SOS;][]{Clemmensen},  $\ell_1$-penalized generalized linear models (logistic regression for binary, and multinomial logistic regression for multi-class problem) \citep[$\ell_1$-GLM;][]{goeman2012penalized}, random forests \citep[RF][]{breiman2001RF,RF}, $\ell_1$-penalized support vector machine \citep[$\ell_1$-SVM;][]{cortes1995support,SVM,bradley1998feature,fung2004feature,becker2009penalizedsvm}. All these methods are designed for vector predictors. From matrix and tensor discriminant analysis literature, we include MDA (matrix discriminant analysis) and PMDA (penalized MDA) from \citet{Zhong2015}, constrained multilinear discriminant analysis (CMDA) and directly generalized tensor discriminant analysis (DGTDA) from \citet{CMDA}, tensor GLM based on CP decomposition \citep[CP-GLM;][]{zhou2013tensor} where rank-3 decomposition has the best performance and is thus reported, and sparse tensor discriminant analysis \citep[STDA][]{STDA}. Most of other methods mentioned in the Introduction Section are not included because they are either unsuitable for multi-class data or too computationally demanding. Furthermore, we also report the error rates of the Bayes' rule as baseline, as well as the oracle vector classifier and oracle tensor classifier that use the oracle information of important predictors. More implementation details can be found in the Supplementary Materials.

In all simulations, we generated the training data such that there are $n_k=75$ observations within each class, unless otherwise specified.  {We generated an independent validation set, where tuning parameters of all methods were chosen with minimum error rates, and an independent testing set for evaluating methods. The validation set has the same number of observations as the training data, while the testing data has $10,000$ observations.} For all the methods, the reported error rates are evaluated on the testing set. All the simulation results are based on 100 replicates of the above procedure. We also compared the variable selection results of the methods quantified by the true positive rate (TPR) and the false positive rate (FPR), defined as:
\begin{equation}\label{selection rate}
\textrm{TPR}=\frac{\mid \widehat{\calD}\bigcap \calD\mid}{\mid \calD\mid},~\textrm{FPR}=\frac{\mid \widehat{\calD}\bigcap \calD^c\mid}{\mid \calD^c\mid}.
\end{equation}
When introducing the models, we use the following shorthand notation. For a matrix $\bolOmega$, $\bolOmega=AR(\rho)$ means $\Omega_{ij}=\rho^{\mid i-j\mid}$ for all $i,j$, while $\bolOmega=CS(\rho)$ means $\Omega_{ij}=\rho$ for $i\ne j$ and $\Omega_{ii}=1$ for all $i$. For a tensor $\mbC$ and a number $c$, $\mbC=c$ means that all the elements in $\mbC$ are equal to $c$. 

\subsection{Models without covariates}\label{subsec:without covs}
We first compare CATCH with existing methods on six models with only tensor predictor but no covariates. The first three models (M1--M3) involve matrix predictors of size $64\times 64$ from $K=4$ classes; the following three (T1--T3) involve $3$-way tensor predictors of size $30\times 36\times 30$ from $K=3$ classes; The last model (T3i) is a special case of T3 with imbalanced (unequal) class sizes. In each model setting, we specify $\bolSigma_m,m=1,\ldots,M$ and $\mbB_k,k=2,\ldots,K$ and set $\bmu_1=0$ and $\bmu_k=\llbracket \mbB_k;\bolSigma_1,\ldots,\bolSigma_M\rrbracket$. Then generate data from TDA model \eqref{TDA}.

We let $\calD_1$ and $\calD_2$ be subsets of $\calD=\calD_1\bigcup\calD_2$ such that $\mbB_{k,\calD^c}=0$ for each $k$. Specifically, $\calD_1=\{(i,j):i=1,2,11,12~\mathrm{and}~j=1,2\}$, $\calD_2=\{(i,j):i=1,2,11,12~\mathrm{and}~j=11,12\}$ and $\calD=\calD_1\bigcup\calD_2$ for models (M1)--(M3); and $\calD_1=\{(i,j,l):i=1,2,11,12,j=1,11~\mathrm{and}~l=1\}$ and $\calD_2=\{(i,j):i=1,2,11,12,j=1,11~\mathrm{and}~l=11\}$ for models (T1)--(T3i).

\noindent {\bf Model (M1) (Independent predictors):} $\bolSigma_1=\bolSigma_2=\mbI_{64}$, $\mbB_{2,\calD}=0.6$, $\mbB_{3,\calD_1}=0.6$, $\mbB_{3,\calD_2}=1.8$, $\mbB_{4,\calD_1}=-0.6$ and $\mbB_{4,\calD_2}=0.6$.

\noindent {\bf Model (M2) (Independent rows):} $\bolSigma_1=\mbI_{64},\bolSigma_2=AR(0.7)$, $\mbB_{2,\calD}=0.4$, $\mbB_{3,\calD_1}=0.4$ and $\mbB_{3,\calD_2}=1.2$, $\mbB_{4,\calD_1}=-0.4$ and $\mbB_{4,\calD_2}=0.4$.

\noindent {\bf Model (M3) (Dependent predictors):} $\bolSigma_1=CS(0.3),\bolSigma_2=AR(0.7)$, $\mbB_{2,\calD}=0.4 $, $\mbB_{3,\calD_1}=0.4$ and $\mbB_{3,\calD_2}=1.2$, $\mbB_{4,\calD_1}=-0.4$ and $\mbB_{4,\calD_2}=0.4$.

\noindent {\bf Model (T1) (Independent predictors):}
$\bolSigma_m,m=1,2,3$ are all identity matrices, $\mbB_{2,\calD}=0.6$, $\mbB_{3,\calD_1}=0.6$ and $\mbB_{3,\calD_2}=1.5$.

\noindent {\bf Model (T2) (Independent mode-2 fibers):}
$\bolSigma_1=AR(0.7),\bolSigma_2=\mbI_{36}, \bolSigma_3=CS(0.3)$, $\mbB_{2,\calD}=0.4$, $\mbB_{3,\calD_1}=0.4$ and $\mbB_{3,\calD_2}=1$.

\noindent {\bf Model (T3) (Dependent predictors):}
$\bolSigma_1=AR(0.7),\bolSigma_2=CS(0.3)$, $\bolSigma_3=CS(0.3)$, $\mbB_{2,\calD}=0.4$, $\mbB_{3,\calD_1}=0.4$ and $\mbB_{3,\calD_2}=1$.

\noindent {\bf Model (T3i) (Imbalanced classes):} 
Same as Model (T3), except for $n_1=n_2=40, n_3=200$.


\begin{table}[H]
	\centering
	\begin{tabular}{c|cccccccc}
		\hline
		Error rate(\%)&M1&M2&M3&T1&T2&T3&T3i&S.E.$\leq$\\
		\hline
		Bayes&14.29&19.24&8.84&14.48&16.17&12.18&8.10&(0.04)\\
		Tensor Oracle&15.71&20.97&9.76&16.28&17.92&13.42&9.21&(0.13)\\
		Vector Oracle&16.20&21.51&10.22&16.56&18.64&14.14&9.48&(0.18)\\\hline
		\textbf{CATCH}&\textbf{17.44}&\textbf{20.09}&\textbf{9.88}&\textbf{19.69}&\textbf{19.05}&\textbf{13.83}&\textbf{9.78}&\textbf{(0.17)}\\
		STDA&74.22&72.69&47.04&66.39&65.00&57.15&35.10&(0.87)\\
		DGTDA&75.06&74.97&52.04&66.74&66.68&60.84&40.67&(0.20)\\
		CMDA&36.27&40.06&19.81&44.6&37.28&27.57&21.75&(0.26)\\		
		MDA&38.34&44.35&29.84&NA&NA&NA&NA&(0.29)\\
		PMDA&27.61&33.55&19.26&NA&NA&NA&NA&(0.46)\\\hline
		$\ell_1$-FDA&18.33&23.98&14.64&30.56&25.12&30.82&25.57&(0.33$\footnotemark$)\\
		SOS&19.21&24.40&11.82&25.88&26.07&20.33&13.09&(0.22)\\
		$\ell_1$-GLM&18.85&22.98&10.85&25.41&22.65&17.22&13.62&(0.16)\\
		RF&53.21&43.75&16.79&NA&NA&NA&NA&(0.17)\\				
		\hline
	\end{tabular}
	\caption{Prediction comparison. The means and the maximum standard errors (in parentheses) of classification error rates are reported. The maximum S.E. of $\ell_1$-FDA has {excluded model (T1), which has S.E. of 1.14. MDA and PMDA are not applicable for $3$-way tensor data; RF can not handle the high-dimensionality in the tensor models; $\ell_1$-SVM and CP-GLM are designed for binary (or pairwise) classifications thus are included only in later binary classification problems. }}
	\label{tab: table1}
\end{table}

\footnotetext{Model T1 has standard error equal to 1.14.}

	\begin{table}[H]
	\centering
	\begin{tabular}{cc|cccccccc}
		\hline
		\multicolumn{1}{c}{ }&&M1&M2&M3&T1&T2&T3&T3i&S.E.$\leq$\\	
		\hline
		\multirow{2}{*}{CATCH}&TPR&99.06&93.94&92.13&83.13&82.13&86.56&71.38&(1.69)\\
		&FPR&0.16&0.12&0.01&0.05&0.03&0.03&0.01&(0.02)\\
		\hline
		\multirow{2}{*}{$\ell_1$-FDA}&TPR&51.75&60.00&52.38&54.38&91.19&93.63&100&(1.82)\\
		&FPR&0.11&0.29&6.88&6.82&1.29&29.28&19.52&(1.02)\\
		\hline
		\multirow{2}{*}{SOS}&TPR&99.31&91.19&82.31&60.19&57.69&60.5&55.94&(1.31)\\
		&FPR&0.38&0.35&0.39&0.05&0.05&0.06&0.08&(0.02)\\
		\hline
		\multirow{2}{*}{$\ell_1$-GLM}&TPR&99.44&92.25&85.00&64.69&65.06&65.69&61.81&(1.01)\\
		&FPR&0.35&0.23&0.23&0.06&0.03&0.03&0.17&(0.05)\\
		\hline
	\end{tabular}
	\caption{Variable selection comparison. TPR and FPR are defined in \eqref{selection rate}.}
	\label{tab: table2}
\end{table} 

The error rates of all methods are reported in Table~\ref{tab: table1}. CATCH significantly outperforms all the other methods, and closely resembles the oracle classifiers across all the models. This supports the application of CATCH. In what follows we discuss the comparison in more details.

First, the comparison among CATCH, MDA and PMDA suggests that it is critical to utilize the sparsity assumption. MDA does not perform variable selection, while PMDA performs variable selection on the rows but not the columns. On the other hand, CATCH can achieve elementwise sparsity. Hence, PMDA significantly improves MDA, while CATCH outperforms PMDA. We can see this point more clearly by noting that the two oracle methods are very close to the Bayes rule, since they have oracle information on the important predictors. By performing variable selection CATCH has accuracy similar to the oracle methods. Moreover, CATCH significantly outperforms other tensor discriminant analysis methods (STDA, DGTDA and CMDA) that more or less based on low dimensional (sparse) projections.

Second, although $\ell_1$-FDA, $\ell_1$-GLM and SOS aggressively take advantage of the sparsity assumption, our CATCH estimator is still more accurate. The margin becomes larger for higher order tensors in Models (T1)--(T3i), and correlated predictors in Models (M3), (T3) and (T3i). This is because CATCH honors the tensor structure and preserves more information. Because $\ell_1$-FDA, $\ell_1$-GLM and SOS require vectorizing data, they are less efficient. The importance of honoring the tensor structure can be further confirmed by examining the two oracle classifiers. The oracle tensor classifier takes into account the tensor structure, and uniformly outperforms the oracle vector classifier.

We also investigated the variable selection results as summarized in Table~\ref{tab: table2}. We did not include other tensor methods in this comparison, because they do not perform variable selection as aggressively as the reported ones. SOS and $\ell_1$-GLM tend to under-select, while $\ell_1$-FDA tends to over-select. CATCH usually selects the majority of the important features, with very few false positives. Such results explain why CATCH performs similarly to the oracle methods. It also supports our theoretical results on the variable selection consistency of CATCH.

{Finally, for the imbalanced classes model (T3i), {the classification is easier than the balanced model (T3), as the Bayes error decreases. Hence, all methods perform better in prediction, but CATCH is still the closest to the Bayes rule. On the other hand, the variable selection becomes more challenging because under-selection would not hurt the prediction much. For example, CATCH is almost as accurate as the oracle estimator when only 71\% of the true variables were selected and near zero false positives. Overall, CATCH is not sensitive to imbalanced classes. }

\subsection{Models with covariates} \label{covariates}
We considered CATCH model \eqref{CATCH_eq1} and \eqref{CATCH_eq2} with tensor predictors of size $30\times 36\times 30$, covariates $\mbU\in \mathbb{R}^2$, and binary classification setting, $K=2$. In each model in the following, we specify CATCH model parameters $\bolphi_k,\bolmu_k,\mbB_k,\bolPsi,\bolSigma_m$ and $\bolalpha=\llbracket \bolalpha^*;\bolSigma_1^{1/2},\ldots,\bolSigma_M^{1/2},\mbI_q\rrbracket$. 
We let $\calD=\{(i,j,l):i=1,2,11,12,j=1,11~\mathrm{and}~l=1,11\}$,  $\bolPsi=\mbI_2$ and two possible values of $\bolalpha^*$ by $\bolalpha^*_1$ and $\bolalpha^*_2$, where $\bolalpha^*_{1,ijst}=0.5$ for $i,j,s=1,\ldots,5$ and $t=1$; $\bolalpha^*_{2,ijst}=1$ for $i,j,s=1,\ldots,15$ and $t=1$; for all other $(k,i,j,s,t)$, $\bolalpha^*_{k,ijst}=0$.

\noindent {\bf Model (C1) (Covariates and tensor with independent predictors):} $\bolSigma_m, m=1,2,3$ are identity matrices, $\mbB_{2,\calD}=0.8$ and $\mbB_{2,\calD^c}=0$,
    $\bolphi_2=0.3, \bolalpha^*=\bolalpha^*_2$.
    
\noindent {\bf Model (C2) (Covariates and tensor with independent mode-2 fibers):}
   $\bolSigma_1=AR(0.7),\bolSigma_2=\mbI_{36}, \bolSigma_3=CS(0.3)$, $\mbB_{2,\calD}=0.4$ and $\mbB_{2,\calD^c}=0$,
    $\bolphi_2=0.3, \bolalpha^*=\bolalpha^*_1$.

\noindent {\bf Model (C3) (Covariates and tensor with dependent predictors):}
 $\bolSigma_1=AR(0.7),\bolSigma_2=CS(0.3), \bolSigma_3=CS(0.3)$, $\mbB_{2,\calD}=0.4$ and $\mbB_{2,\calD^c}=0$,
    $\bolphi_2=0.3, \bolalpha^*=\bolalpha^*_1$.

\noindent {\bf Model (C3a) (Independent covariates and tensor):} Replace $\bolphi_2=1$ and $\bolalpha^*=\mathbf{0}$ in (C3).
    
\noindent {\bf Model (C3b) (Non-discriminantive covariates):} Same as (C3), except for $\bolphi_2=\mathbf{0}$.

\noindent {\bf Model (C3i) (Imbalanced classes):} Same as (C3), except for $n_1=40$ and $n_2=200$.

To investigate how each methods can incorporate covariates information in practice, we considered two tasks for each methods: classification based on $\mbX$ alone or based on both $\mbX$ and $\mbU$. In the presence of $\mbU$, only CATCH and CP-GLM \citep{zhou2013tensor} can naturally include the covariates in their model. Thus for other tensor methods (STDA, DGTDA and CMDA), we stack covariates with downsized tensor when these methods apply nearest neighbor to classify; for vector methods, we stack $\mbU$ and $\vecc(\mbX)$ as a single vector. {As suggested by a referee, we also included two methods, SOS weighted and $\ell_1$-GLM weighted that first fit models based on $\mbX$ and $\mbU$ separately and then combine the information with a weighted vote. In SOS weighted, we apply LDA on $(Y,\mbU)$ and SOS on $(Y,\vecc(\mbX))$ to obtain $\widehat\bolalpha\T\mbU$ and $\widehat\bolbeta\T\vecc(\mbX)$, respectively. Then we find a weighted vote of them by applying LDA to $Y$ with $\{\widehat\bolalpha\T\mbU, \widehat\bolbeta\T\vecc(\mbX)\}$ as predictors. Similarly, in $\ell_1$-GLM weighted, $\mbU$ and $\vecc(\mbX)$ are first separately modeled by GLM and $\ell_1$-GLM, respectively and then combined together with GLM.}

The results are listed in Table~\ref{tab: table5}. CATCH again uniformly outperforms all the competitors. Moreover, when the covariates are included, the performance of CATCH can be significantly improved. On the other hand, including the covariates in other methods in general does not improve classification performance, except for model (C3a) where the covariate and the tensor are independent within each class. {This suggests that adjusting covariates is important and that simply embedding the two part information without any study on their relationship is not efficient enough.}  We further study the role of covariates in models (C3a, b, i) as follows. 

Model (C3a) is a special case where $\mbU$ and $\mbX$ are independent within classes since $\bolalpha=0$. Both the covariates and the tensor predictors contribute to the classification, but LDA and CATCH without covariates are less accurate than CATCH, because they do not utilize both types of predictors. Naively combining $\mbX$ (or $\vecc(X)$) and $\mbU$ under this scenario can actually improve many methods (CP-GLM, CMDA, SOS, $\ell_1$-GLM and $\ell_1$-SVM) significantly, but CATCH is still superior to them.

Model (C3b) is another special case where covariates affect the tensor predictors but themselves do not contribute to the classification since $\bolphi_1=\bolphi_2=0$. It can be seen that the LDA has an error rate around $50\%$ since the covariates do not have any power in the classification. Since the tensor predictors are very informative, CATCH without covariates has already achieved a very error rate comparing to other methods. Still, when we adjust for the covariates in CATCH, we have a significant improvement. This reinforces our point that even when covariates are not important themselves, they should still be included for further analysis. Without adjusting for the tensor regression relationship between the tensor and the covariates, all other methods cannot effectively utilize the additional information from $\mbU$ and thus fail to improve.

In Model (C3i), the Bayes error is lower than that in Model (C3). Consequently, all the methods have improved accuracy, but CATCH remains the best classifier.

Finally, the variable selection results also show that CATCH outperforms all the competitors and inclusion of the covariates leads to better variable selection. These results can be found in the Supplementary Materials.

\begin{table}[t]
	\centering
	\begin{tabular}{ccccccccc}
		\hline
		\multicolumn{2}{c}{Error rate(\%)}&C1&C2&C3&C3a&C3b&C3i&S.E. $\leq$\\
		\hline
		\multicolumn{2}{c}{Bayes}&5.33&10.97&8.15&6.08&8.39&5.45&(0.03)\\
		\hline
		LDA&$\mbU$&42.62&42.21&42.45&24.34&50.02&16.72&(0.18)\\
		\hline
		\multirow{2}{*}{CATCH}&$\mbX$&31.03&21.30&16.7&10.78&14.76&10.58&(0.59)\\
		&$\mbX,\mbU$&\textbf{11.12}&\textbf{16.67}&\textbf{11.24}&\textbf{8.33}&\textbf{11.28}&\textbf{7.36}&\textbf{(0.22)}\\
		
		\hline
		\multirow{2}{*}{CP-GLM}&$\mbX$&40.54&27.45&17.93&16.15&18.02&10.75&(1.40)\\
		&${\mbX,\mbU}$&39.40&31.16&19.65&13.82&19.12&10.59&(1.30)\\
		
		\hline
		\multirow{2}{*}{STDA}&$\mbX$&48.31&46.17&44.69&41.45&46.42&23.54&(0.38)\\
		&${\mbX,\mbU}$&48.2&46.08&44.65&40.11&46.54&23.62&(0.34)\\
		\hline
		\multirow{2}{*}{DGTDA}&$\mbX$&49.7&49.42&44.99&44.08&46.84&26.01&(0.16)\\
		&$\mbX,\mbU$&49.55&49.68&45.88&45.65&47.68&26.73&(0.17)\\
		\hline
		\multirow{2}{*}{CMDA}&$\mbX$&39.70&34.60&27.86&25.59&29.16&16.03&(0.32)\\
		&$\mbX,\mbU$&39.70&34.45&27.27&22.5&28.73&15.81&(0.32)\\

		\hline
		\multirow{2}{*}{$\ell_1$-FDA}&$\vecc(\mbX)$&43.36&30.29&31.72&20.73&30.45&27.32&(0.65)\\
		
		&$\vecc(\mbX),\mbU$&43.37&30.31&31.71&20.56&33.36&27.30&(0.65)\\
		\hline
		\multirow{3}{*}{SOS}&$\vecc(\mbX)$&34.30&17.46&14.39&11.87&15.56&9.29&(0.31)\\
		
		&$\vecc(\mbX),\mbU$&34.30&17.45&14.39&9.70&15.56&9.29&(0.31)\\
		&Weighted&33.26&16.97&13.78&9.08&14.61&8.91&(0.29)\\

		\hline
		\multirow{3}{*}{$\ell_1$-GLM}&$\vecc(\mbX)$&34.01&17.57&14.58&12.31&15.87&10.43&(0.30)\\
		
		&$\vecc(\mbX,\mbU)$&34.07&17.8&14.95&9.78&15.90&10.43&(0.35)\\
		&Weighted&33.11&17.8&14.55&9.74&15.4&9.51&(0.35)\\
		\hline  
		
		\multirow{2}{*}{$\ell_1$-SVM}& $\vecc(\mbX)$&25.53&25.94&19.05&16.85&19.00&10.14&(0.20)\\
		
		&$\vecc(\mbX,\mbU)$&23.71&22.91&19.03&15.3&18.21&10.95&(0.10)\\
		%
		\hline  
	\end{tabular}
	\caption{Prediction comparison. The means and the maximum standard errors (in parentheses) of classification error rates based on 100 replicates are reported. Same as models (T1)--(T3i) in Table~\ref{tab: table1}, MDA, PMDA and RF are excluded because MDA and PMDA are not applicable for 3-way tensors and RF can not handle the high-dimensionality.}
	\label{tab: table5}
\end{table}


To demonstrate the applicability of CATCH in high-dimensional data, we further considered variants of models (C1)--(C3) where we increase the tensor dimensions to $80\times 80\times 80$, that is $512,000$ voxels in total. Many methods become practically inapplicable because of the prohibitive computational costs. Therefore, we only compare CATCH with $\ell_1$-GLM, $\ell_1$-FDA, CP-GLM and DGTDA. The classification errors and the variable selection results can be found in the Supplementary Materials, where CATCH continues to achieve better accuracy and variable selection than the competitors.

We also compared the computational costs for CATCH and the competitors. Because CATCH contains two steps of adjusting for covariates and penalized estimation of the coefficients $\mbB_k$, we report the computation time for these two steps along with the total of them. While CATCH can produce the whole solution path simultaneously, many methods cannot. Therefore, we only compare the methods for pre-chosen tuning parameters that yield the highest accuracy for each method, respectively. For discriminant analysis methods, we need to first find the means and the covariances. This step can be sped up easily by parallel computing and is hence excluded when we calculate the computation time. The average computation time from 20 replicates for model (C3) and its higher-dimension variation is listed in Table~\ref{Time}. It can be seen that the total computation time for CATCH is shorter than most methods, except for $\ell_1$-FDA and $\ell_1$-GLM. This shows that CATCH is a computationally efficient method in general. For the comparison of CATCH $\ell_1$-FDA and $\ell_1$-GLM, we note that the fast computation of $\ell_1$-FDA is somewhat expected, because it assumes that the covariance is diagonal (and we did not include the computational time for the standardization step in $\ell_1$-FDA that centers and standardizes each variable). This simplification greatly improves the computational speed. However, this assumption may lead to lower classification accuracy, as seen in the numerical studies. On the other hand, the penalized estimation step of CATCH has similar computational cost as $\ell_1$-GLM  because both methods use coordinate descent methods. The major difference between CATCH and $\ell_1$-GLM comes from the part where we adjust for the covariates. But we have seen that this step repay us with considerable classification accuracy. Meanwhile, this step can be finished much faster if we implement it in a parallel fashion as the adjustment is element-wise. Hence, the added computational cost of adjusting for the covariates should not be a serious issue. It is also worth mentioning that CP-GLM \citep{zhou2013tensor} is the best existing tensor method we found in the literature, in terms of both accuracy (Table~\ref{tab: table5}) and speed (Table~\ref{Time}). Nonetheless, CATCH substantially improves both classification accuracy and computational speed. Moreover, under the higher dimension $80\times80\times80$, CP-GLM requires a warm-start from first downsize the tensor to a smaller size and obtain an initial estimator. Even a rank-1 CP-GLM model has more than $240$ model parameters, which is more than the sample size. While our CATCH model fitting requires no warm-start and is more feasible to high-dimensional sparse situations. In Table~\ref{Time}, we have included the warming-up stage of rank-3 CP-GLM. If we use rank-1 CP-GLM, the classification error will be much worse (than the results in Table~\ref{tab: table5}), while the computational time is reduced to $12.86$ seconds but is still longer than CATCH's $8.04$ seconds.

\begin{table}[t]
	\centering
	\begin{tabular}{ccccccccc}
		\hline
		Dimension&\multicolumn{3}{c}{CATCH}&$\ell_1$-FDA&SOS&CP-GLM\\
		&Adjust&Estimation&Total&\\
		\hline
		$30\times 36\times 30$&0.13&0.17&0.3&0.06&2.79&1.62\\
		$80\times 80\times 80$&3.38&4.66&8.04&1.27&70.79&18.34\\
		\hline
		&STDA&CMDA&DGTDA&$\ell_1$-GLM&$\ell_1$-SVM\\
		\hline
		$30\times 36\times 30$&4.63&109.36&1.96&0.19&29.49\\
		$80\times 80\times 80$&22.59&NA&51.16&2.86&NA\\
		\hline
		
	\end{tabular}
	\caption{Computation time (seconds) averaged from 20 replicates. The parameters were chosen as in Model (C3) (dimension $30\times 36\times 30$) and Model (C3H) (in Supplementary Materials, dimension $80\times 80\times 80$). The method $\ell_1$-SVM and CMDA failed to converge in Model (C3H) in an hour and hence their computation time is reported as ``NA''. }\label{Time}
\end{table}

\section{Real data analysis}\label{sec:realdata}

In this section, we apply CATCH to a colorimetric sensor array data with matrix predictors $\mbX$, and a neuroimaging application with 3-way tensor predictors $\mbX$ and covariates to diagnose the attention deficit hyperactivity disorder (ADHD). The analysis of another dataset on diagnosing autism (the ASD dataset) is presented in the Supplementary Materials. The analysis on the ADHD and the ASD datasets lead to similar conclusions from the statistical perspective, so we only include one of them in the main body of our paper. 

%
%
\subsection{The Colorimetric Sensor Array Data}

Colorimetric sensor arrays (CSA) are devices that identify volatile chemical toxicants (VCT). They use chemical dyes to turn the smell of a chemical to optical composite signals. This results in $36\times 3$ matrix predictors, where each row contains the color change of a dye before and after exposure, and the three columns correspond to red, green and blue, respectively.

The CSA data were collected on $n=147$ chemicals to classify them into $K=21$ classes.  One class is non toxic chemical, while the other 20 classes are high hazard toxic industrial chemicals. The CSA are exposed to the chemicals at two conditions: the Immediately Dangerous to Life or Health (IDLH) concentrations for 2 minutes, and the Permissible Exposure Level (PEL) for 5 minutes. The CSA data was used in \citep{Zhong2015} to demonstrate MDA and PMDA. Following their approach, we analyze the two conditions separately.

We applied CATCH, $\ell_1$-FDA, MDA, PMDA, SOS, $\ell_1$-multinomial, Random Forest, SVM and STDA to the IDLH and the PEL datasets. In each replicate, we randomly sampled 21 observations as the testing set and used the rest 126 observations as the training set. We used $K-1=20$ discriminant directions for $\ell_1$-FDA, MDA and PMDA. These methods allow users to choose the number of discriminant directions, but we observe that cross validation over this parameter leads to minimal improvement of performance.  The other tuning parameters are chosen by 5-fold cross validation on the training set.

The classification error rates are listed in Table~\ref{tab: colorimetric}. At the IDLH level, all methods have excellent accuracy. In particular, CATCH, $\ell_1$-FDA and SOS achieve perfect classification. Meanwhile, when the CSAs are exposed to chemicals at the PEL level, the classification becomes much more difficult, possibly because of the low concentration of chemicals. Random forest is the best classifier, while CATCH is the second best that significantly outperforms all the other methods. However, CATCH also has some noticeable advantages over random forest. First, CATCH performs variable selection to allow easy interpretation. Second, CATCH can handle much higher dimensions, while random forest would not be applicable, such as in the ADHD dataset in Section~\ref{sec:ADHD}. Third, the classifier fitted by random forest is difficult to interpret, while CATCH provides a low-dimensional representation of the data, as we now discuss.

\begin{table}[t!]
	\centering
	\begin{tabular}{c|ccccccc}
		\hline
		Error (\%)&CATCH&STDA&DGTDA&CMDA&MDA&PMDA\\
		IDLH&0 (0)&1.7 (0.3)&0.2 (0.1)&0.4 (0.1)&1.6 (0.1)&2.4 (0.1)\\
		PEL&3.2 (0.1)&11.2 (0.6)&7.2 (0.4)&5.1 (0.4)&18.9 (0.2) &19.7 (0.1)\\
		\hline
		&LDA&$\ell_1$-FDA&SOS&$\ell_1$-GLM&RF& SVM\\
		IDLH&5.1 (0.3)&0 (0)&0 (0)&0.6 (0.2)&0.2 (0.1)&0.7 (0.2)\\
		PEL&20.1 (0.9) &5.7 (0.1)&10.7 (0.1)&17.5 (0.6)&1.7 (0.3)&4.9 (0.10)\\
		\hline
	\end{tabular}
	\caption{Colorimetric sensor array data analysis. Classification error rates of colorimetric sensor array data under IDLH and PEL exposure conditions. Mean and standard error (in parentheses) of error rates of 100 replicates are recorded.}
	 \label{tab: colorimetric}
\end{table}

To visualize the classification results of CATCH on IDLH case, we performed principal component analysis on $\langle\hatmbB_2,\mbX\rangle,\ldots,\langle \hatmbB_{20},\mbX\rangle$, where $\hatmbB_k$ are given by the CATCH estimator. Since the first two principal components explained over $90\%$ of the total variability of the 20 discriminative components, we plotted these two principal components in Figure \ref{fig:colorimetric}. It can be seen that the different classes fall into different clusters, with very little overlap.

\begin{figure}[t!]
\centering
\includegraphics[width=0.8\textwidth]{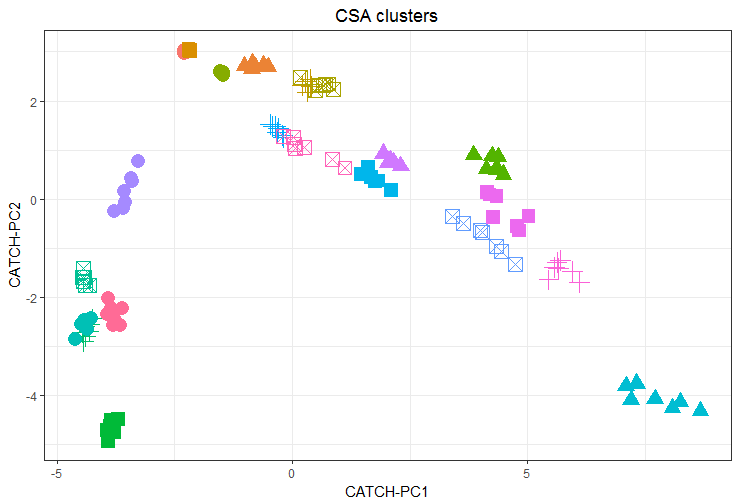}
\caption{Colorimetric sensor array data projected on first two principal component directions of $\langle \widehat{\mbB}_k,\mbX\rangle, k=2,\ldots,K$ in the IDLH experiment. There are 147 observations from 21 class in total.}
\label{fig:colorimetric}
\end{figure}

\begin{table}[t!]
	\centering
	\begin{tabular}{c|cc|cc|cc|cc}
		\hline
		\multirow{3}{*}{Scenarios}&\multicolumn{4}{c}{$30\times36\times30$}&\multicolumn{4}{c}{$24\times27\times24$}\\
		&\multicolumn{2}{c}{Binary}&\multicolumn{2}{c}{Multiclass}&\multicolumn{2}{c}{Binary}&\multicolumn{2}{c}{Multiclass}\\
	
		&Mean&\multicolumn{1}{c}{SE}&Mean&\multicolumn{1}{c}{SE}&Mean&\multicolumn{1}{c}{SE}&Mean&\multicolumn{1}{c}{SE}\\
		\hline
		\textbf{CATCH}&\textbf{23.57}&\textbf{0.2}&\textbf{36.11}&\textbf{0.23}&\textbf{22.79}&\textbf{0.24}&\textbf{35.22}&\textbf{0.25}\\
		CP-GLM&25.19&0.24&NA&NA&25.05&0.19&NA&NA\\
		STDA&32.44&0.35&49.35&0.33&31.01&0.29&49.35&0.28\\
		DGTDA&29.88&0.30&46.41&0.31&30.51&0.29&47.17&0.30\\
		CMDA&30.3&0.28&46.29&0.32&30.38&0.26&47.94&0.3\\
		SOS&24.11&0.28&37.48&0.26&23.87&0.26&37.07&0.29\\
		Weighted SOS &24.12&0.11&38.89&0.17&24.35&0.10&38.47&0.19\\
		$\ell_1$-GLM&23.75&0.17&35.42&0.22&23.99&0.16&35.66&0.21\\
		Weighted $\ell_1$-GLM&24.32&0.27&37.81&0.28&23.31&0.22&37.34&0.28\\
		$\ell_1$-SVM&26.95&0.3&40.79&0.29&27.54&0.31&41.28&0.32\\
		\hline
	\end{tabular}
	\caption{Classification errors on the ADHD datasets of two different tensor sizes. Testing classification error rates, standard errors are based on 100 replicates of training/testing sets. CP-GLM is not applicable for multiclass problem because multinomial logistic is not available.}
		\label{Resize}
\end{table}

\subsection{The ADHD dataset}\label{sec:ADHD}
We further considered the attention deficit hyperactivity disorder (ADHD) data set, which contains both tensor predictors and covariates. Neuro Bureau shares the ADHD dataset on NITRC (\url{http://fcon\_1000.projects.nitrc.org/indi/adhd200}) \citep{ADHD}. It contains complete rs-fMRI and s-MRI data for 930 individuals, along with their age, gender and handedness. The T1-weighted MRI are downsized to $30\times 36\times 30$ in our analysis. {We further downsize the tensors to $24\times 27\times 24$ and compared the results side-by-side with the $30\times 36\times 30$ tensor data.} These individuals fall into four categories: Typically Developing Children (TDC), ADHD Combined, ADHD Hyperactive and ADHD Inattentive. The covariate gender is binary. We stratify the datasets by fitting CATCH on male and female subjects separately. After stratification on gender, we have two continuous covariates, age and handedness. Then we pool error rates from the two subsets to measure the performance of CATCH. MDA and PMDA cannot be applied to this dataset because the images are three-way tensors rather than matrices. The $\ell_1$-FDA is not included, because it seems to be overly sensitive to tuning parameters on this dataset. 

 We split the data into a training set of 762 subjects and a testing set of 168 subjects. We tested the performance of CATCH in two classification problems. Because only 13 subjects have ADHD Hyperactive, we combine them with the ADHD Combined class. This gives us a three-class problem. Further, because subjects with ADHD combined and ADHD hyperactive have symptoms of hyperactivity, while subjects do not in the other two categories, TDC and ADHD Inattentative, we group them into two classes. This results in a binary problem. 

Before we fit classifiers on this datasets, we replace zero tensor element by half minimum nonzero elements and perform the log transformation $\log(\mbX)$ such that the variables are on the same scale and more normally distributed. The classification results are listed in Table~\ref{Resize}. Overall, CATCH has the best performance among all methods.

\section{Discussion}\label{sec:discussion}

In this paper, we develop the CATCH model and construct an accurate classifier when both tensor and covariates are present. We give an intensive study on how to integrate the information from the tensor and the covariates through both direct and indirect effects. The superior performance of the proposed method is demonstrated through both theoretical and numerical studies. Although we only considered low-dimensional continuous covariates, the CATCH model framework can be extended in the future to accommodate applications where some of the covariates are discrete, and to imaging genetics applications with high-dimensional covariates. 

{
In the CATCH model, we assume that the covariates and the adjusted tensor predictors are normal with constant covariance across classes. In the future, it will be interesting to study how to relax these model assumptions. One possible direction is to relax the constant covariance assumption. Such a development can be viewed as parallel to the extension from LDA to quadratic discriminant analysis (QDA), although we have a much more complicated problem. In the literature, several authors have studied how to perform sparse QDA for vector data \citep{fan2015quadro,li2015sparse,jiang2015quda, le2014sparse}. These results are likely to facilitate our future research.} 

Another important direction for future research is to relax the normality assumption. As pointed out by the associate editor, transformations are often helpful in relaxing normality assumptions. \citet{lin2003discriminant,han2013coda,SeSDA} discussed methods to transform the data such that they satisfy the discriminant analysis type of assumptions. It will be interesting to investigate the integration of their techniques with CATCH to relax the normality assumption. We leave this topic as future research.

\baselineskip=5pt

\bibliographystyle{agsm}
\bibliography{draft}

\end{document}